\documentclass[12pt]{article}

\usepackage{graphicx}
\usepackage{amsfonts}
\usepackage{amssymb}
\usepackage{amsmath}
\usepackage[colorlinks,urlcolor=blue]{hyperref}

\begin{document}

\title{\bf Quaternions in Hamiltonian dynamics of a rigid body -- Part I }
\providecommand{\keywords}[1]{\textbf{\textit{Keywords: }} #1}

\author{
  Stanislav S. Zub\\
  Faculty of Cybernetics,\\
  Taras Shevchenko National University of Kyiv,\\
  Glushkov boul., 2, corps 6.,\\
  Kyiv, Ukraine 03680\\
  \texttt{stah@univ.kiev.ua}\\
  \\
  Sergiy I. Zub\\
  Institute of Metrology,\\
  Mironositskaya st., 42,\\
  Kharkiv, Ukraine 61002\\
  \texttt{sergii.zub@gmail.com}}

\maketitle

\newpage
\begin{abstract}
This paper showed that Poisson brackets in quaternion variables can be obtained directly 
from canonical Poisson brackets on cotangent bundle of $SE(3)$ (or $SO(3)$) 
endowed by canonical symplectic geometry.

Quaternion parameters in our case are just dynamic variables 
in canonical Hamiltonian mechanics of a rigid body on $T^*SE(3)$

The obtained results based on quaternions representation as explicit functions 
of rotation matrix elements of $SO(3)$ group.

The relation of obtained Poisson structure to the canonical Poisson 
and symplectic structures on $T^*S^3$ were investigated.

To derive the motion equations of Hamiltonian dynamics in quaternionic variables 
it is proposed to use the mixed frame of reference where 
translational degrees of freedom describes in the inertial frame of reference and 
degree of rotational freedom in the body frame.

It turns out that motion equations for system with Hamiltonian of the rather general form 
can be written in algebraic operations on quaternions.

.

\end{abstract}

\newpage
\tableofcontents

%
\newcommand{\s}[1]{\ensuremath{\boldsymbol{\sigma}_{#1}}}
\newcommand{\bsym}[1]{\ensuremath{\boldsymbol{#1}}}
\newcommand{\w}{\ensuremath{\boldsymbol{\wedge}}}
\newcommand{\lc}{\ensuremath{\boldsymbol{\rfloor}}}
\newcommand{\rc}{\ensuremath{\boldsymbol{\lfloor}}}

\thispagestyle{empty}

\newpage
\section{Introduction}
\label{intro}

This article is describing main results of Hamiltonian dynamics of a rigid body in quaternion variables.

There are many of publications that are devoted to application of quaternions in mechanics of a rigid body. 
The most of them refer to kinematics of a rigid body, i.e. description of rigid body orientation 
in quaternion parameters~\cite{Whittaker:NYork44}.

Hereinafter we specify those rare articles that were devoted to description of dynamics with using quaternions.
Kozlov's work \cite{Kozlov:Igevsk95} gives rigid body dynamics in Lagrangian representation 
based on classic approach developed by Poincare. 
Borisov and Mamaev \cite{BM:NonLinePS97,BM:PS99,BM:RBD01} proceeding from the deep relations between 
quaternion algebra and groups $SO(3),SO(4)$ have proposed Poisson brackets between 
parameters of quaternion and components of intrinsic angular momentum of a rigid body 
and considering them as some generators of a Lie-Poisson structure. 
Therewith condition of normalization of quaternion per unit that 
is required for a rigid body description emerges as the special value 
of the Casimir function in that Lie-Poisson structure.

Expression of matrix element of the group in quaternion parameters is well known \cite{Whittaker:NYork44,Marsden:IMS94,BM:PS99,BM:RBD01} 
that is also clearly demonstrates structure of orthogonal matrices. 
Per se inverse problem of expressing quaternion parameters 
in terms of elements of the corresponding rotation matrix is not so difficult.
But in literature we have found (post factum) only one article devoted to solution of that task
\cite{Salamin:Stanford79}.  
Main goal of this short but contensive article is the comparative analysis of  
necessary number of operations for computation through matrix or quaternion.

In terms of quantity was proven that quaternions computation is more effective. 
Using of quaternion also has advantage of stability 
for numerical integration of motion equations (see \cite[p.104]{BM:PS99}).

From general relations given in work \cite{Zub:LeeGOPM13} 
one can deduce following expressions for Poisson brackets between 
elements of rotation matrix and components of intrinsic angular momentum 
in inertial frame of reference.

Here and further for determinacy will consider $SE(3)$ 
that describes not only rotational 
but also translational degrees of freedom of a rigid body.
\[
\begin{cases}
   \{x_i, x_j\} = 0; \quad \ \ \{p_i, p_j\} = 0; \quad \ \ \{m_i, Q_{jk}\} = \varepsilon_{ijl}Q_{lk}; \\
   \{x_i, p_j\} = \delta_{ij}; \quad \{p_i, Q_{jk}\} = 0; \quad \{m_i, m_j\} = \varepsilon_{ijl}m_l,\\
   \{x_i, Q_{jk}\} = 0; \quad \{p_i, m_j\} = 0; \\
   \{x_i, m_j\} = 0; \quad \ \{Q_{ij}, Q_{kl}\} = 0; \\
\end{cases}
\leqno(1)\]
where
$x_i$ -- coordinates of the body center of mass,
$p_i$ -- momentum components of translational motion,
$Q_{jk}$ -- elements of rotation matrix that describe the body orientation with respect to inertial frame of reference,
$m_j$ -- components of intrinsic angular moment of the body with respect to inertial frame of reference.

In fact expression (1) will also be proved during the explanation by the different ways.

From the explicit expression of quaternion parameters as functions of elements 
of rotation matrix from (1) one can deduce required Poisson brackets 
between components of quaternion and angular momentum components of the body. 
Obtaining of these relations is main goal of the paper.

It turns out that expressions obtained in this way 
have the same form that previously discussed Lie-Poisson brackets 
between generators of Poisson structure mentioned above (see \cite[(2.7), p. 103]{BM:PS99}).
 
But obtained relations have other core idea.
Quaternion parameters in our case are not generators of Poisson structure, 
but dynamic variables in canonical Hamiltonian mechanics on $T^*SE(3)$ (it is not Lie-Poisson structure). 
In our case structural tensor of Poisson brackets of Lie algebra is not nondegenerate, 
and, so, Casimir functions are absent and normalization condition for quaternion  
just reflect relations between quaternion dynamic variables on group.
To illustrate that fact let us consider example.
Let $\varphi$ is one of Euler angel. 
Then dynamic variable $\rm cos^2(\varphi)+sin^2(\varphi)$ is identically equal to one, 
but it is not the Casimir function.

After obtaining Poisson structure for quaternion variables as dynamic variables 
for a rigid body (on $T^*SO(3)$) we investigate its relation with canonical Poisson 
and symplectic structures on $T^*S^3$ (the relation is local isomorphism).

To derive equations of motion in Hamiltonian dynamics 
in quaternion variables is proposed to use a mixed frame of reference 
where translational degrees of freedom describe in inertial frame of reference, 
and the rotational in the body frame
(a reference frame associated with a body). 
It turns out that motion equations can be written through algebraic operations in quaternions.

Symplectic and Poisson structures on $T^*S^3$ can be set completely independently from algebraic structure of quaternion variables,
but one can see that Poisson brackets for these dynamic variables is completely determine laws of multiplication in quaternion algebra.

Hamiltonian dynamics of a rigid body in quaternion variables 
is an interesting point of intersection of two pioneer directions in mathematics 
that were offered by Hamilton: 
powerful stream of Hamiltonian formalism that covering almost all areas of mathematical physics 
and not so strong but surprisingly beautiful stream of quaternionic applications 
in various fields of mathematics and physics.

\newpage
\section{ Hamiltonian formalism on $T^*SO(3)$}
\label{Symplectic}

\bigskip
\subsection{ Representation of right trivialization of $T^*SO(3)$}

Let's bring are some useful relations for group $SO(3)$
and its cotangent bundle many of which can be found in \cite{AbrMar02,MarRat98}.

Group $SO(3)$ is formed from orthogonal unimodular matrices $\mathbf{R}$
i.e. $\mathbf{R}^T = \mathbf{R}^{-1}$, $\rm det(\mathbf{R})= 1$.
Accordingly the Lie algebra $\bsym{so(3)}$ is formed from antisymmetric $3\times 3$-matrices 
with a Lie bracket in form of matrix commutator.
Let's introduce isomorphism of vector spaces $\hat{}:\mathbb{R}^3\rightarrow\bsym{so(3)}$
such that \cite[p. 285]{MarRat98}:
$\widehat{\bsym{\xi}}_{kl} = -\varepsilon_{ikl}\xi_i$,
$\xi_i = -\frac12 \varepsilon_{ikl} \widehat{\bsym{\xi}}_{kl}$,
where $\varepsilon_{ikl}$ --- Levi-Civita symbol. Then
\begin{equation}
\begin{cases}
   \widehat{\bsym{\xi}}\bsym{\eta} = \bsym{\xi}\times \bsym{\eta}; \quad
   [\widehat{\bsym{\xi}},\widehat{\bsym{\eta}}] =
   \widehat{\bsym{\xi}}\widehat{\bsym{\eta}} - \widehat{\bsym{\eta}}\widehat{\bsym{\xi}} =
   \widehat{\bsym{\xi}\times \bsym{\eta}};\\
   \langle \bsym{\xi},\bsym{\eta} \rangle
   = -\frac12 {\rm tr}(\widehat{\bsym{\xi}}\widehat{\bsym{\eta}}); \quad
   \mathbf{B}\widehat{\bsym{\xi}}\mathbf{B}^{-1} = \widehat{\mathbf{B}\bsym{\xi}}
\end{cases}
\label{eq1}
\end{equation}
The scalar product above allows to set equivalence of
a Lie algebra and its dual space $\bsym{so(3)}^*\simeq\bsym{so(3)}$.
Symbol ``$\simeq$'' mean a diffeomorphism. 
As a rule, diffeomorphisms used below have a simple group-theoretical 
or differential-geometrical meaning that is explained in cited literature.

The representation of {\it right trivialization}
corresponds to {\it inertial frame of reference}
that is shown in \cite[p. 314]{AbrMar02}, we have \\
$T SO(3)\simeq SO(3)\times\bsym{so(3)}$, 
$T^*SO(3)\simeq SO(3)\times\bsym{so(3)}^*$.

Then right and left group action of $SO(3)$ on $T^*SO(3)$ (Cotangent Lift \cite[p. 166]{MarRat98})
takes the form
\begin{equation}
\begin{cases}
      R^{ct}_{\mathbf{B}}: (\mathbf{R},\widehat{\bsym{\pi}})\in T^*SO(3)
      \rightarrow(\mathbf{R}\mathbf{B},\widehat{\bsym{\pi}}), \quad \mathbf{B}\in SO(3);\\
      L^{ct}_{\mathbf{B}}: (\mathbf{R},\widehat{\bsym{\pi}})\in T^*SO(3)
      \rightarrow(\mathbf{B}\mathbf{R},\mathbf{B}\widehat{\bsym{\pi}}\mathbf{B}^{-1})
      = (\mathbf{B}\mathbf{R},{\rm Ad}^*_{\mathbf{B}^{-1}}\widehat{\bsym{\pi}})
\end{cases}
\label{eq2}
\end{equation}

\newpage
\subsection{ Symplectic and Poisson structures on $T^*SO(3)$ }

The Liouville 1-form on $T^*SO(3)\simeq SO(3)\times\bsym{so(3)}^*$ is given by:
\begin{equation}
\Theta^{T^*SO(3)}_{|(\mathbf{R},\bsym{\pi})}
  = -\frac12 {\rm tr}(\widehat{\bsym{\pi}}\widehat{\bsym{\delta R}})
  = \pi_i\delta R^i,
\label{eq3}
\end{equation}
where $\widehat{\bsym{\delta R}} = (d\mathbf{R})\mathbf{R}^{-1}$ --- right-invariant of the Maurer-Cartan 1-form.

Then for canonical symplectic 2-form,
using Maurer-Cartan equation \cite[ё. 276]{MarRat98} we have:
\begin{equation}
\Omega^{T^*SO(3)}_{can}  = -d\Theta^{T^*SO(3)} = {\delta R}^i\w d\pi_i - \pi_i [\delta R,\delta R]^i =
\label{eq4}
\end{equation}
\[ = -\frac12\varepsilon_{ijk}\delta R_{jk}\w d \pi_i
   + \frac12 \pi_i\varepsilon_{ijk}\delta R_{js}\w \delta R_{sk}
\]

Any given symplectic structure $\Omega$ defines a Poisson structure on the same manifold:
\begin{equation}
\{F, G\}(z) = \Omega(\xi_F(z),\xi_G(z))
  = \partial_{\xi_G}F = -\partial_{\xi_F}G
\label{eq5}
\end{equation}
where for vector field $\xi_G$ the relation $i_{\xi_G}\Omega = dG$ is fulfilled.

Let's consider the elements of the matrix $\mathbf{R}$ and components of the moment $\bsym{\pi}$
as the dynamic variables on $T^*SO(3)$ and we obtain the following set of Poisson brackets
that completely determine Poisson structure on $T^*SO(3)$:
\begin{equation}
\{R_{ij}, R_{kl}\} = 0, \quad
\{\pi_i, R_{jk}\} = \varepsilon_{ijl}R_{lk}, \quad
\{\pi_i, \pi_j\} = \varepsilon_{ijl}\pi_l.
\label{eq6}
\end{equation}

Notice that in inertial frame of reference Poisson brackets for matrix elements
$\mathbf{R}$ is grouped together in columns, for example,
Poisson brackets of 3-rd column expressed only through elements of 3-rd column.

\newpage
\section{ Quaternion algebra }
\label{Algebra}

Quaternions form an associative algebra with unit \( e_0 \) and generators \\ 
\( e_i \), $i=1,2,3$ that satisfy to defining relation 
\[ e_r e_s = -\delta_{rs} e_0 + \varepsilon_{rst} e_t
\eqno(1)
\]
or, that is equivalent to
\[
\begin{cases}
   [e_r, e_s] = e_r e_s - e_s e_r = 2\varepsilon_{rst} e_t; \\
   e_r e_s + e_s e_r = -2\delta_{rs} e_0.
\end{cases}
\eqno(2)
\]
Thus quaternions form a 4-dimensional vector space over the field of real numbers
\[ q = q^0 e_0 + q^1 e_1 + q^2 e_2 + q^3 e_3
\eqno(3) \]
or, through components of a 4-dimensional column vector: $q = (q^0, q^1, q^2, q^3)$.

The component $q^0$ is a scalar part of a quaternion $q$
and components $q^1, q^2, q^3$ grouped as a vector part.
Thus quaternion can be written as $q = (q^0, \bsym{q})$.

If \( q^0 = 0 \) then permissible form \( q=\bsym{q} \),
and such quaternion called pure quaternion \cite[p. 301]{Marsden:IMS94}.
The pure quaternions form a linear subspace of quaternions algebra,
but not subalgebra, 
because associative product of two pure quaternions leads to quaternion of general form

From (1) follows law of quaternion multiplication $a$ and $b$
\[ a b = (a^0 e_0 + a^r e_r)(b^0 e_0 + b^s e_s)
       = (a^0 e_0 + \bsym{a})(b^0 e_0 + \bsym{b})
\eqno(4) \]
\[ = (a^0 b^0 - \langle\bsym{a},\bsym{b}\rangle) e_0 + a^0 \bsym{b} + b^0 \bsym{a} + \bsym{a}\times \bsym{b} \]

Then for pure quaternions through law of associative multiplication we have
the following expressions of scalar and vector product.
\[
\begin{cases}
   \langle\bsym{x},\bsym{y}\rangle = -\frac12(\bsym{x}\bsym{y} + \bsym{y}\bsym{x}); \\
   \bsym{x}\times\bsym{y} = \frac12(\bsym{x}\bsym{y} - \bsym{y}\bsym{x})
\end{cases}
\eqno(5)\]

Operation of quaternions conjugation is introduced as follows
\[ e_0{}^\dag = e_0,\qquad e_k{}^\dag  = -e_k
\eqno(6) \]
or
\[ q^\dag = (q^0, \bsym{q})^\dag = (q^0, -\bsym{q})
\eqno(7) \]

Thus pure quaternions is completely characterized by the following property
\[ \bsym{x}^\dag = -\bsym{x}
\eqno(8) \]

From (7) follows
\[ q q^\dag = q^\dag q = (q^0 q^0 + (\bsym{q},\bsym{q})) e_0 \\
   = (q^0)^2 + (q^1)^2 + (q^2)^2 + (q^3)^2
\eqno(9) \]

Thus, from (4,7) follows
\[ (a b)^\dag = b^\dag a^\dag
\eqno(10) \]

Let introduce norm of quaternion
\[  |q| = \sqrt{(q q^\dag)}
\eqno(11) \]
then
\[ |ab| = |a||b|
\eqno(12) \]

In addition from (9,11) follow a simple representation of inverse quaternion
\[ q^{-1} = \frac{q^\dag}{|q|^2}
\eqno(13) \]

Formulas (9) and (13) indicate that all quaternions but excluding a zeroth have its own inverse,
i.e. algebra of quaternions is the division ring.

From (12) and (13) that a unit quaternions,
i.e. quaternions that norms are equal to unit form a group.

\newpage
\section{ Presentation of rotations in quaternions }
\label{Rotation1}

Let's consider linear subspace of pure quaternions as 3-dimension
Euclidean space with inner product that determined by formula (5) \S3.

Let's show that inner automorphism of quaternion algebra
that generated by unit quaternion $q$, translate space of pure quaternions into itself.

The law of transformation can be written as
\[ \bsym{x}^{'} = q\bsym{x}q^{-1} = q\bsym{x}q^{\dag} = Q[\bsym{x}],\quad |q| = 1.
\leqno(1)
\]
Let's show that resulting quaternion $\bsym{x}^{'}$ also is a pure quaternion, indeed,
\[ (q\bsym{x}q^{\dag})^{\dag} = (q^{\dag})^{\dag}\bsym{x}^{\dag}q^{\dag} = -q\bsym{x}q^{\dag}.
\leqno(2)
\]
so, operator $Q[\bsym{x}]$ is a linear operator
that acting in a subspace of pure quaternions.

Let's show that operator $Q[\bsym{x}]$ preserves scalar product of vectors,
using expression of scalar product through associative multiplication from (5) \S3
\[ \langle Q[\bsym{x}],Q[\bsym{y}]\rangle = -\frac12(q\bsym{x}q^{-1}q\bsym{y}q^{-1}+q\bsym{y}q^{-1}q\bsym{x}q^{-1})
= -\frac12 q(\bsym{x}\bsym{y}+\bsym{y}\bsym{x})q^{-1}
\]
\[
= q\langle\bsym{x},\bsym{y}\rangle q^{-1} = \langle\bsym{x},\bsym{y}\rangle.
\]
Thus, following relations for invariance of operator $Q$ are correct (proved similarly)
\[
\begin{cases}
   \langle Q[\bsym{x}],Q[\bsym{y}]\rangle = \langle\bsym{x},\bsym{y}\rangle; \\
   Q[\bsym{x}]\times Q[\bsym{y}] = Q[\bsym{x}\times \bsym{y}]; \\
   \langle Q[\bsym{z}],Q[\bsym{x}]\times Q[\bsym{y}]\rangle
   = \langle \bsym{z},\bsym{x} \times \bsym{y} \rangle
\end{cases}
\leqno(3)\]

The first relation (3) means that operator $Q$ is orthogonal
and third relation that it is unimodular, i.e. $Q$ is proper rotation ($Q~\in~SO (3)$).

The particular interest in context of posed a problem 
is explicit form of matrix elements of operator $Q$ 
that can be obtained from (1) by using multiplication law (4) \S3
\[ Q_{ik} = 2 \left[\left((q^0)^2 - \frac12\right)\delta_{ik}\\
    + q^i q^k - q^0 q^j\varepsilon_{jik}\right]
\leqno(4)\]
that for unit quaternion is equivalent to following expression for matrix of operator
\[ Q =  \begin{bmatrix}
           q_0^2+q_1^2-q_2^2-q_3^2 & 2(q_1 q_2 - q_0 q_3) & 2(q_1 q_3 + q_0 q_2)  \\
           2(q_1 q_2 + q_0 q_3) & q_0^2-q_1^2+q_2^2- q_3^2 & 2(q_2 q_3 - q_0 q_1) \\
           2(q_1 q_3 - q_0 q_2) & 2(q_2 q_3 + q_0 q_1) & q_0^2-q_1^2-q_2^2+q_3^2
        \end{bmatrix}
\leqno(5) \]

Let's show that $q \longrightarrow Q$ is mapping on whole $SO(3)$ group.

In particular case when $q_1=0, q_2=0$
this matrix takes the form
\[ Q =  \begin{bmatrix}
           q_0^2-q_3^2 & -2 q_0 q_3 & 0  \\
           2 q_0 q_3 & q_0^2 - q_3^2 & 0  \\
           0 & 0 & q_0^2+q_3^2
        \end{bmatrix}
\]
If $q_0=\cos{\frac12\theta},\quad q_3=\sin{\frac12\theta}$ then
\[ Q =  \begin{bmatrix}
           \cos{\theta} & -\sin{\theta} & 0  \\
           \sin{\theta} & \cos{\theta} & 0  \\
           0 & 0 & 1
        \end{bmatrix}
\]
This is a matrix of rotation around $Z$ axis 
through an angle $\theta$ (counterclockwise).

Likewise, we will get rotations around $x$ and $y$ axes.

{\bf Notice}. It is useful to note that an arbitrary unit quaternion can be represented as
\[ q = \rm \cos{(\varphi/2)} e_0 + \sin{(\varphi/2)} \bsym{e},
\leqno(6)\]
where $\bsym{e}$ is pure and unit quaternion
that sets the rotation axis and $\varphi$ 
is corresponding angle of rotation around that axis.
Expression (6) that is considered as a function $\varphi$
is one-parameter subgroup of the group of unit quaternions.

Because of $\Gamma:q \longrightarrow Q$ is a homomorphism, 
so product of such rotations be contained in the image of this homomorphism.
It is known that any rotation can be obtained as a product 
of rotations around the axes of Cartesian coordinate system.
So, matrices (4) and (5) are record of an arbitrary element of the group $SO(3)$ 
in terms of quaternion parameters.

Thus formula (1) defines a homomorphism of the group of unit quaternions on $SO(3)$.
At the same time, these groups are locally isomorphic and the group of unit quaternions 
is double covering group $SO(3)$ \cite{Marsden:IMS94}.
Indeed, from (1) it is obvious that quaternion $q$ and $(-q)$ gives the same rotation $Q$.

\newpage
\section{ Quaternion representation in terms of rotations }
\label{Rotation2}

In previous section it was shown that every rotation $SO(3)$
corresponds to 2 and only 2 unit quaternions of the different sign.
There is the problem of explicit expression of this functional dependence.

Matrix $Q$ that corresponds to quaternion $q$ has following elements
\[ Q_{ik} = (2q_0^2 - 1)\delta_{ik}
    + 2 q_i q_k - 2 q_0 q_j\varepsilon_{jik}
\leqno(1)\]

Let's obtain the trace of matrix with taking into account
that $q$ is a unit quaternion
\[ \rm Sp(Q) = 4 q_0^2 - 1
\leqno(2)\]
i.e.
\[ q_0^2 = \frac14(\rm Sp(Q) + 1)
\leqno(3)\]

The antisymmetric part of matrix $Q$ has a simple form, so
\[ q_0 q_i = -\frac14 \varepsilon_{ijk}Q_{jk}
\leqno(4)\]

Thus where $q_0\neq 0$ one can explicitly express components of quaternion
corresponding to a given rotation matrix, through its elements
\[
\begin{cases}
   q_0 = \frac12(\rm Sp(Q) + 1)^{\frac12}; \\
   q_i = -\frac12\frac{\varepsilon_{ijk}Q_{jk}}{(\rm Sp(Q) + 1)^{\frac12}} ;
\end{cases}
\leqno(5)\]

There are two solutions to this system,
corresponds to two various choices of sign of root
in the expression for $q_0$.

If we want to have a solution of these functional equations
in the neighborhood of quaternion with zero scalar part $(\rm Sp(Q)=-1)$,
i.e. in neighbourhood of pure quaternion then formulas (5) are not suitable.

It should be also mentioned that for numerical computations
the difficulties can also arise for small but non-zero values of $q^0$.
Therefore E. Salamin who investigated the problem of comparative evaluation
of effectiveness of numerical computations with using orthogonal matrices
and quaternions \cite{Salamin:Stanford79}, suggested to use next set of formulas
to find components of quaternion that follows from matrix $Q$
\[
\begin{cases}
   q_0^2 = \frac14(1 + Q_{11} + Q_{22} + Q_{33}); \\
   q_1^2 = \frac14(1 + Q_{11} - Q_{22} - Q_{33}); \\
   q_2^2 = \frac14(1 - Q_{11} + Q_{22} - Q_{33}); \\
   q_3^2 = \frac14(1 - Q_{11} - Q_{22} + Q_{33}); \\
\end{cases}
\leqno(6)\]
\[
\begin{cases}
   q_0 q_1 = \frac14(Q_{32} - Q_{23}); \\
   q_0 q_2 = \frac14(Q_{13} - Q_{31}); \\
   q_0 q_3 = \frac14(Q_{21} - Q_{12}); \\
   q_1 q_2 = \frac14(Q_{12} + Q_{21}); \\
   q_1 q_3 = \frac14(Q_{13} + Q_{31}); \\
   q_2 q_3 = \frac14(Q_{23} + Q_{32}); \\
\end{cases}
\leqno(7)\]

For numerical computations it is optimal to choose of the maximum component
of quaternion (by its absolute value) from (6) 
(choice of the sign of component will determine the sign of quaternion), 
and then the others components can be found from (7).

It is clear that for unit quaternion at least one of the component will be different from 0.

Note that solution (5) that found above corresponds to 1-st line of (6) and first 3 lines in (7).

\newpage
\section{ Poisson brackets with components of quaternion}
\label{Poisson1}

As it is follows from the foregoing description quaternion variables
can be regarded as dynamic variables in the phase space of a rigid body.
The only difference from usual dynamic variables in two-valuedness of quaternion variables.

This is not an obstacle to use of quaternion variables
to describe of dynamics of a rigid body.
By selecting a certain sign of these variables at the initial point of the trajectory 
in phase space we choose the sign in accord with the continuity throughout the trajectory.

Also, if we restrict some neighborhood of a point of phase space of a rigid body
then it is possible to choose one of branch of expressions in quaternion variables
express in terms of elements of rotation matrix.

To derive Poisson brackets with components of quaternions
we need to compute $\frac{\partial q_0}{\partial Q_{kl}}$
and $\frac{\partial q_i}{\partial Q_{kl}}$.

For instance,
\[\{\pi_i,q_0\} = \frac{\partial q_0}{\partial Q_{kl}}\{\pi_i,Q_{kl}\}
\leqno(1)\]
where Poisson bracket $\{\pi_i,Q_{kl}\}$ is known from (1) \S1.

From (5) of previous section we have
\[
\begin{cases}
   \frac{\partial q_0}{\partial Q_{kl}} = \frac1{8 q_0}\delta_{kl}; \\
   \frac{\partial q_j}{\partial Q_{kl}}
  = -\frac1{4 q_0}\left(\varepsilon_{jkl} + \frac12\frac{q_j}{q_0}\delta_{kl}\right);
\end{cases}
\leqno(2)\]

If we take into account $\{\pi_i,Q_{kl}\}=\varepsilon_{ikn}Q_{nl}$ with (2) we get
\[\{\pi_i,q_0\} = \frac{\partial q_0}{\partial Q_{kl}}\{\pi_i,Q_{kl}\}
  = \frac12 q_i
\leqno(3)\]
and
\[\{\pi_i,q_j\} = \frac12(\varepsilon_{ijk}q_k - q_0\delta_{ij})
\leqno(4)\]

Since $\{Q_{ij},Q_{kl}\}=0$ we finally obtain
\[
\begin{cases}
   \{q_\mu,q_\nu\} = 0, \quad \mu = 0,1,2,3; \\
   \{\pi_i,q_0\} = \frac12 q_i ; \\
   \{\pi_i,q_j\} = \frac12(\varepsilon_{ijk}q_k - q_0\delta_{ij});\\
\end{cases}
\leqno(5)\]

As mentioned in the introduction,
such Poisson brackets were obtained by Borisov and Mamaev
as an expressions for generators of some Poisson structure of Lie-Poisson.
In addition the value of
\(C(q)=(q^0)^2 + (q^1)^2 + (q^2)^2 + (q^3)^2\)
is Casimir function in Poisson structure and so in order to get description 
of dynamics of a rigid body we should pass on 
to a symplectic leaf of this Poisson structure that corresponds to the value $C(q) =1$.

\bigskip
\subsection{ The properties of invariance of Poisson structure on $T^*(SE(3))$ ($T^*(SO(3))$) }

Take note on Poisson brackets (1) \S1 that contain matrix elements of rotation.
For example, look at expression of
\[ \{\pi_i, Q_{jk}\} = \varepsilon_{ijl}Q_{lk}
\leqno(1)\]
letТs multiply it on fixed matrix $B\in SO(3)$ from right side.
Since elements of the matrix  $B$ are constants
then we can put them into Poisson brackets in the left side of (1)
then the expression is valid
\[ \{\pi_i, (Q B)_{jn}\} = \varepsilon_{ijl}(Q B)_{ln}
\leqno(2)\]

If we performs the same conversion with all such Poisson brackets
we can see that matrix elements of rotation $P=QB $ will satisfies
to all those Poisson brackets as matrix elements of initial rotation $Q$.

Let's show that Poisson brackets of previous section
can be written in a more compact form
by using law of quaternion multiplication.

Let's multiply quaternion $q=q_0 e_0 + q_k e_k$ from left side on $e_i$
\[e_i q = q_0 e_i + q_k (e_i e_k)
        = q_0 e_i + q_k (-\delta_{ik}e_0 + \varepsilon_{ikj} e_j)
        = - q_i e_0 + q_0\delta_{ij}e_j + q_k\varepsilon_{ikj} e_j
\]
i.e.
\[e_i q = - q_i e_0 + q_0\delta_{ij}e_j + q_k\varepsilon_{ikj} e_j
\leqno(3)\]
or
\[
\begin{cases}
   (e_i q)_0 = - q_i ; \\
   (e_i q)_j = -\varepsilon_{ijk}q_k + q_0\delta_{ij};\\
\end{cases}
\leqno(4)\]

Then
\[
\begin{cases}
   \{q_0, \pi_i\} = \frac12 (e_i q)_0 ; \\
   \{q_j, \pi_i\} = \frac12 (e_i q)_j;\\
\end{cases}
\leqno(5)\]
i.e.
\[  \{q_\mu, \pi_i\} = \frac12 (e_i q)_\mu, \quad \mu = 0,1,2,3;\\
\leqno(6)\]

Formally one can multiply (6) from left side
on fixed quaternions $e_0,e_1,e_2,e_3$
and then summing by $\mu$-index.
In addition using $e_\mu$ as constant
we can insert them into Poisson bracket.

The result is
\[  \{\pi_i, q\} = -\frac12 e_i q;
\leqno(7)\]

In the same way as for matrices when we multiply (7) from the right side
on a some fixed quaternion $b,|b|=1$
we can see that new quaternion dynamic variable $p=q b$
satisfies same Poisson brackets (7) like $q$.
In fact it is a simple algebraic transformation
is equivalent to applying matrices of right translation (2) \S2
to the set of relationships (7).

Therefore $p$ satisfies to all Poisson brackets
that are analogous to Poisson brackets in (5) of previous section.

\bigskip
\subsection{ Poisson brackets with quaternion components, the special case }

Strictly speaking the obtaining relations (5) \S6
that are present above will not be correct
in the neighbourhood of pure quaternion
$\rm Sp(Q)=-1\longrightarrow q_0=0$,
because (5) \S5 is not feasible the solution at this point.

But this not means that formulas (5) \S6
in the neighbourhood of pure quaternion is not valid.
A quaternion with $q_0=0$ does not looks like a critical point for these formulas.
However the proof of these formulas in general requires other approaches.

First of all one can use SalaminТs solutions of (6,7) \S5
by taking as a pivotal the component of quaternion
that is not zero in this neighborhood, for example  $q^1$.
Unfortunately such direct approach leads to very cumbersome and non-transparent computations.

Let's consider properties of invariance
that were listed in previous section, as well as the fact
that the map $q\longrightarrow Q$
is a homomorphism on group $SO(3)$
as it was described above.

Let's suppose that we need to obtain Poisson bracket
with quaternions in the neighbourhood of pure quaternion $q$ with $q_0=0$.

LetТs consider such fixed quaternion $b:|b|=1$
that dynamic quaternion variable $p=q b$
has value $p^0\neq 0$ at this point
(this always can be done for $q:|q|=1$ as it was shown above).
Quaternion $p$ mapping into the matrix $P\in SO(3)$
for the chosen homomorphism ($p \longrightarrow P$).

Quaternion variable $p$ has same relation to matrix-variable $P$
just as the relation of variable $q$ to variable $Q$ when deriving
Poisson brackets with quaternions that was described above and moreover $p^0\neq 0$.

Consequently for quaternion variable $p$
Poisson brackets (5) \S6 are fulfilled.
By applying the inverse transform $q=p b^{-1}$
to the obtained formulas we will got (5) \S6
but for quaternion variable $q$.

\newpage
\section{ Tangent and cotangent bundle over $S^3$ }
\label{Diff}

\bigskip
\subsection{ Tangent bundle to group $S^3\subset\mathbb{H}$ }

Since group $S^3$ is a subset of associative algebra $\mathbb{H}$ (so and linear space)
then a curve on the group is also a curve in the linear space.
Therefore the vector of speed $\xi$ of this curve i.e. element $\xi\in T(S^3)$
we can considered as the element of $\mathbb{H}$.
\[ q(t) q(t)^{\dag} = 1\longrightarrow
   \dot{q} q^{\dag} + q \dot{q}^{\dag} =
   \langle q, \dot{q}\rangle = 0
\leqno(1)\]

So, tangent vectors to $S^3$ in the point $q$
can be presented as quaternions $\xi$
that are orthogonal to quaternion $q$.

The left (right) translation of tangent vector
in this presentation coincides with product
of quaternion that are tangent vector
on quaternion that generates left (right) translation.

Really, let $q(t)_{|t=0}=a$
\[ T_a L_b\cdot\dot{q} = \frac{d}{d t} (L_b q(t))_{|t=0}
 = \frac{d}{d t} (b q(t))_{|t=0} = b\dot{q}
\leqno(2)\]

{\bf Proposition 1}. If $\bsym{e}$ --- pure quaternion
then $\xi=L_q\bsym{e}=q\bsym{e}\in T(S^3)$ and, vice versa,
if $\xi\in T(S^3)$
then $q^{-1}\xi=q^\dag\xi$ --- pure quaternion ($q\in S^3$).

Similar statements are correct and for right translations.

$\square$

The fact that $\bsym{e}$ -- pure quaternion can be express in terms of
\[ \bsym{e} + \bsym{e}^{\dag} = 0\longrightarrow
   \bsym{e}e_0^\dag + e_0\bsym{e}^{\dag} = 0\longrightarrow
   \langle e_0, \bsym{e}\rangle = 0 \longrightarrow
   \langle L_q e_0, L_q\bsym{e}\rangle = 0
\]
and, vice versa
\[ \langle q, \xi\rangle = 0\longrightarrow
   \langle L_{q^{-1}}q, L_{q^{-1}}\xi\rangle = 0\longrightarrow
   \langle e_0, q^{-1}\xi\rangle = 0\longrightarrow
   q^{-1}\xi + (q^{-1}\xi)^\dag = 0
\]

$\blacksquare$

{\bf Conclusion}. The tangent space at the group identity 
can be identified with subspace of pure quaternions.

\bigskip
On $T(S^3)$ from $\mathbb{H}$ the Euclidean metric is induced.
In particular $T_e(S^3)$ is a subspace of pure quaternions
and at the same time it is 3-dimensional Euclidean space.

\bigskip
\subsection{ The Lie algebra of $S^3\subset\mathbb{H}$ }

As shown above an arbitrary unit quaternion can be represented as
\( \rm \cos{(\varphi/2)} e_0 + \sin{(\varphi/2)} \bsym{e} \),
where $\bsym{e}$ --  pure and unit quaternion.

Chosen the one-half angle as a parameter of the one-parameter subgroup
connected with the interpretation of homomorphism
$\Gamma: S^3\mapsto SO(3)$ wherein $\bsym{e}$ defines an axis of rotation
and $\varphi$ -- corresponding the angle of rotation around this axis.

For study of the Lie algebra $Lie(S^3)$
will be more convenient to present one-parameter subgroup in form
\[  {\rm exp}\left(t\bsym{e}\right)
  = {\rm \cos}{(t)} e_0 + {\rm \sin}{(t)} \bsym{e}
\leqno(1)\]

Accordingly the pure unit quaternions $\bsym{e}$
can be considered as an element of $Lie(S^3)$ algebra.

It is reasonable to assume as a basis for $Lie(S^3)$
set of generators \( e_i \), $i=1,2,3$.

Then formula (2) \S3 provides structure constants of this Lie algebra
\[ [e_r, e_s] = e_r e_s - e_s e_r = 2\varepsilon_{rst} e_t
 = c^t{}_{r s} e_t
\leqno(2) \]

So, from the above it follows that $Lie(S^3)=T_e(S^3)$ can be identified with subspace of pure quaternions.

{\bf Proposition 2}. $Lie(S^3)=T_e(S^3)$ subspace is revealed an Euclidean space where operator $Ad_q\in SO(3)$.

$\square$

Indeed,
\[ {\rm Ad}_q[\bsym{e}] = \frac{d}{d t} (q{\rm exp}(t\bsym{e})q^{-1})_{|t=0}
 = q\bsym{e}q^{-1} = q\bsym{e}q^{\dag} = \Gamma(q)[\bsym{e}]
\leqno(3)\]

$\blacksquare$

Now it is easy to derive expression for Lie bracket in quaternions
\[ {\rm ad}_{\bsym{\xi}}[\bsym{\eta}]
 =  \frac{d}{d t} {\rm Ad}_{{\rm exp}(t\bsym{\xi})}[\bsym{\eta}]_{|t=0}
 = \frac{d}{d t}\left( {\rm exp}(t\bsym{\xi})\bsym{\eta}{\rm exp}(-t\bsym{\xi})\right)
 = \bsym{\xi}\bsym{\eta} - \bsym{\eta}\bsym{\xi}
\leqno(3)\]
\[ {\rm ad}_{\bsym{\xi}}[\bsym{\eta}]
 = \bsym{\xi}\bsym{\eta} - \bsym{\eta}\bsym{\xi}
 = [\bsym{\xi},\bsym{\eta}] = 2\bsym{\xi}\times\bsym{\eta}
\leqno(3a)\]

\bigskip
\subsection{ The dual space to Lie algebra and bundle $T^*(S^3)$ }

The metric on $Lie(S^3)=T_e(S^3)$ allows us to identify $Lie^*(S^3)$ and $Lie(S^3)$.

Indeed, pure quaternion $\bsym{\kappa}$
defines linear form on $Lie(S^3)=T_e(S^3)$
by the formula
\[ \bsym{\kappa}[\bsym{\xi}]
 = \langle\bsym{\kappa},\bsym{\xi}\rangle,\quad
 \forall\bsym{\xi}\in Lie(S^3)
\leqno(1)\]

Let's compute the operator ${\rm Ad}^*_q[\bsym{\mu}]$
\[ \langle{\rm Ad}^*_q[\bsym{\mu}], \bsym{\xi}\rangle
 = \langle\bsym{\mu}, {\rm Ad}_q[\bsym{\xi}]\rangle
 = \langle\bsym{\mu}, q\bsym{\xi}q^{-1}\rangle
 = \langle q^{-1}\bsym{\mu}q, \bsym{\xi}\rangle
 = \langle q^{\dag}\bsym{\mu}q, \bsym{\xi}\rangle
\leqno(2)\]
or
\[{\rm Ad}^*_q[\bsym{\mu}]
 = q^{-1}\bsym{\mu}q = q^{\dag}\bsym{\mu}q
\leqno(2a)\]

Operator ${\rm ad}^*_{\bsym{\xi}}$
can be found either as derivative with respect to $t$ for ${\rm Ad}^*_q[\bsym{\mu}]$
where $q(t)={\rm exp}(t\bsym{\xi})$
or as operator adjoint to ${\rm ad}_{\bsym{\xi}}$.

In any case we get the following result
\[{\rm ad}^*_{\bsym{\xi}}[\bsym{\mu}]
 = [\bsym{\mu}, \bsym{\xi}] = 2\bsym{\mu}\times\bsym{\xi}
\leqno(3)\]

In the light of above said
and bearing in mind the existence of 2-directional invariant metrics
on $S^3$ we can identify $T^*(S^3)$ and $T(S^3)$.

\bigskip
\subsection{ Algebra isomorphism $Lie(S^3)$ and $Lie(SO(3))$ }

Let's consider expression of right-invariant Maurer-Cartan 1-form
that was introduced in subsection 2.2. and represent it through quaternion variables.

\bigskip
{\bf Proposition 3}. The following formulas are valid.

\[ (d Q Q^{-1})_{i k}  = -2 \left(q_i d q_k - q_k d q_i
    + \left(q_0 d q_r - q_r d q_0\right) \varepsilon_{ikr}\right)
\leqno(1)\]

\[ (d Q Q^{-1})_{i} = -\frac12\varepsilon_{irs}(d Q Q^{-1})_{rs}
\leqno(2)\]
\[  = 2\left(q_0 d q_i - q_i d q_0
    + \varepsilon_{irs}q_r d q_s\right) = 2 (d q q^{-1})_i
\]

$\square$

We have
\[ Q_{ij} = (2q_0^2 - 1)\delta_{ij}
    + 2 q_i q_j - 2 q_0 q_r\varepsilon_{ijr}
\leqno(3)\]
\[ d Q_{ij} = 4 q_0 d q_0\delta_{ij}
           + 2 q_i d q_j + 2 q_j d q_i
           - 2 q_0 d q_r\varepsilon_{ijr}  - 2 q_r d q_0 \varepsilon_{ijr}
\leqno(4)\]
\[ Q^{-1}_{j k} = Q^{T}_{j k} = (2q_0^2 - 1)\delta_{j k}
    + 2 q_j q_k + 2 q_0 q_s\varepsilon_{jks}
\leqno(5)\]

From here by using $q_0^2+\bsym{q}^2=1\longrightarrow q_0 d q_0 + \langle\bsym{q},d\bsym{q}\rangle=0$
\[  (d Q Q^{-1})_{i k}
    = \left(4 q_0 d q_0\delta_{ij}
           + 2 q_i d q_j + 2 q_j d q_i
           - 2 q_0 d q_r\varepsilon_{ijr}  - 2 q_r d q_0 \varepsilon_{ijr}\right)
\]
\[  \times\left( (2q_0^2 - 1)\delta_{j k}
    + 2 q_j q_k + 2 q_0 q_s\varepsilon_{jks}  \right)
\]
\[  = 4 (2q_0^2 - 1)q_0 d q_0\delta_{ik}
    + 2 (2q_0^2 - 1) q_i d q_k + 2 (2q_0^2 - 1)q_k d q_i
    - 2 (2q_0^2 - 1) q_0 d q_r\varepsilon_{ikr}  - 2 (2q_0^2 - 1)q_r d q_0 \varepsilon_{ikr}
\]
\[  + 8 q_0 d q_0 q_i q_k
    + 4 q_i d q_j q_j q_k + 4 \bsym{q}^2 q_k d q_i
    - 4 q_0 q_j q_k  d q_r\varepsilon_{ijr}  - 4 q_j q_k q_r d q_0 \varepsilon_{ijr}
\]
\[  + 8 q_0 q_s q_0 d q_0\varepsilon_{iks}
           + 4 q_0 q_i q_s\varepsilon_{jks} d q_j
           - 4 q_0^2 q_s\varepsilon_{jks}\varepsilon_{ijr} d q_r
           - 4 q_r  q_s\varepsilon_{jks}\varepsilon_{ijr} q_0 d q_0
\]
\[  = 4 (2q_0^2 - 1)q_0 d q_0\delta_{ik}
    + 2 (2q_0^2 - 1) q_i d q_k + 2 (2q_0^2 - 1)q_k d q_i
    - 2 (2q_0^2 - 1) q_0 d q_r\varepsilon_{ikr}  - 2 (2q_0^2 - 1)q_r d q_0 \varepsilon_{ikr}
\]
\[  + 8 q_0 d q_0 q_i q_k
    + 4 q_i q_k q_j d q_j + 4 \bsym{q}^2 q_k d q_i
    - 4 q_0 q_j q_k  d q_r\varepsilon_{ijr}  - 4 q_j q_k q_r d q_0 \varepsilon_{ijr}
\]
\[  + 8 q_0^2 q_s d q_0\varepsilon_{iks}
    + 4 q_0 q_i q_s\varepsilon_{jks} d q_j
\]
\[  + 4 q_0^2 q_s (\delta_{ik}\delta_{rs} - \delta_{kr}\delta_{is}) d q_r
    + 4 q_r  q_s(\delta_{ik}\delta_{rs} - \delta_{kr}\delta_{is}) q_0 d q_0
\]
\[  = 4 (2q_0^2 - 1)q_0 d q_0\delta_{ik}
    + 2 (2q_0^2 - 1) q_i d q_k + 2 (2q_0^2 - 1)q_k d q_i
\]
\[  - 2 (2q_0^2 - 1) q_0 d q_r\varepsilon_{ikr}  - 2 (2q_0^2 - 1)q_r d q_0 \varepsilon_{ikr}
\]
\[  + 8 q_i q_k q_0 d q_0
    + 4 q_i q_k q_r d q_r + 4 \bsym{q}^2 q_k d q_i
    - 4 q_0 q_j q_k  d q_r\varepsilon_{ijr}  - 4 q_j q_k q_r d q_0 \varepsilon_{ijr}
\]
\[  + 8 q_0^2 q_s d q_0\varepsilon_{iks}
    + 4 q_0 q_i q_s\varepsilon_{jks} d q_j
\]
\[  + 4 \left(q_0^2 q_r d q_r + \bsym{q}^2 q_0 d q_0\right) \delta_{ik}
    - 4 q_k  q_i q_0 d q_0  - 4 q_0^2 q_i d q_k
\]
\[  = 4\left( (2q_0^2 - 1)q_0 d q_0 + \left(q_0^2 q_r d q_r + \bsym{q}^2 q_0 d q_0\right)\right) \delta_{ik}
\]
\[ + 2 \left((2q_0^2 - 1)   - 2 q_0^2\right) q_i d q_k
   + 2 \left((2q_0^2 - 1)  + 2 \bsym{q}^2\right) q_k d q_i
\]
\[  - 2 (2q_0^2 - 1) q_0 d q_r\varepsilon_{ikr}  - 2 (2q_0^2 - 1)q_r d q_0 \varepsilon_{ikr}
\]
\[  + 4 q_i q_k q_0 d q_0
    + 4 q_i q_k q_r d q_r
    - 4 q_0 q_j q_k  d q_r\varepsilon_{ijr}
\]
\[  + 8 q_0^2 q_s d q_0\varepsilon_{iks}
    + 4 q_0 q_i q_s\varepsilon_{jks} d q_j
\]

\newpage
\[  = 4\left( q_0^2 (q_0 d q_0  +  q_r d q_r) + (q_0^2 + \bsym{q}^2) q_0 d q_0 - q_0 d q_0\right) \delta_{ik}
\]
\[ + 2 \left((2q_0^2 - 1)   - 2 q_0^2\right) q_i d q_k
   + 2 \left((2q_0^2 - 1)  + 2 \bsym{q}^2\right) q_k d q_i
\]
\[  - 2 (2q_0^2 - 1) q_0 d q_r\varepsilon_{ikr}  - 2 (2q_0^2 - 1)q_r d q_0 \varepsilon_{ikr}
\]
\[  + 4 q_i q_k q_0 d q_0
    + 4 q_i q_k q_r d q_r
    - 4 q_0 q_j q_k  d q_r\varepsilon_{ijr}
\]
\[  + 8 q_0^2 q_s d q_0\varepsilon_{iks}
    + 4 q_0 q_i q_s\varepsilon_{jks} d q_j
\]
\[ = -2 q_i d q_k + 2 q_k d q_i
\]
\[  - 4 q_0^2 q_0 d q_r\varepsilon_{ikr} + 2 q_0 d q_r\varepsilon_{ikr}
     + 2 q_r d q_0 \varepsilon_{ikr}
\]
\[  + 4 q_0^2 q_r d q_0 \varepsilon_{ikr}
    + 4 q_0 q_i q_r\varepsilon_{jkr} d q_j
    - 4 q_0 q_j q_k  d q_r\varepsilon_{ijr}
\]
\[ = -2 q_i d q_k + 2 q_k d q_i
   - 4 q_0^2 (q_0 d q_r - q_r d q_0) \varepsilon_{ikr}
\]
\[   + 2 q_0 d q_r\varepsilon_{ikr}
     + 2 q_r d q_0 \varepsilon_{ikr}
     + 4 q_0 q_j(q_i \varepsilon_{kjr} - q_k \varepsilon_{ijr}) d q_r
\]

Let's use the identity
\[ q_i \varepsilon_{kjr} - q_k \varepsilon_{ijr}
   = \frac12\varepsilon_{iks}\varepsilon_{smn}(q_m \varepsilon_{njr} - q_n \varepsilon_{mjr})
\]
\[ = \frac12\varepsilon_{iks}(q_m \varepsilon_{njr}\varepsilon_{smn} - q_n \varepsilon_{mjr}\varepsilon_{smn})
\]
\[ = \frac12\varepsilon_{iks}
     (q_m(\delta_{js}\delta_{mr} - \delta_{rs}\delta_{jm})
     + q_n(\delta_{js}\delta_{nmr} - \delta_{rs}\delta_{jn})
\]
\[ = \varepsilon_{iks}(q_r\delta_{js} - q_j\delta_{rs})
\]

Then
\[  (d Q Q^{-1})_{i k}
    = -2 q_i d q_k + 2 q_k d q_i
    - 4 q_0^2 (q_0 d q_r - q_r d q_0) \varepsilon_{ikr}
\]
\[   + 2 q_0 d q_r\varepsilon_{ikr}
     + 2 q_r d q_0 \varepsilon_{ikr}
     + 4 q_0 q_j\varepsilon_{iks}(q_r\delta_{js} - q_j\delta_{rs}) d q_r
\]
\[  = -2 q_i d q_k + 2 q_k d q_i
    - 4 q_0^2 (q_0 d q_r - q_r d q_0) \varepsilon_{ikr}
\]
\[   + 2 q_0 d q_r\varepsilon_{ikr}
     + 2 q_r d q_0 \varepsilon_{ikr}
     + 4 q_0 \varepsilon_{iks}( q_s q_r d q_r - \bsym{q}^2 d q_s)
\]
\[  = -2 q_i d q_k + 2 q_k d q_i
\]
\[  - 4 q_0 (q_0^2 d q_r + \bsym{q}^2 d q_r - q_r q_0 d q_0 + q_r q_0 d q_0) \varepsilon_{ikr}
\]
\[   + 2 q_0 d q_r\varepsilon_{ikr}
     + 2 q_r d q_0 \varepsilon_{ikr}
\]
\[  = -2 q_i d q_k + 2 q_k d q_i
    - 2 q_0 d q_r \varepsilon_{ikr}
    + 2 q_r d q_0 \varepsilon_{ikr}
\]

\newpage
Thus, it is proved the relation (1), i.e.
\[ (d Q Q^{-1})_{i k}  = -2 \left(q_i d q_k - q_k d q_i
    + \left(q_0 d q_r - q_r d q_0\right) \varepsilon_{ikr}\right)
\]

Going to the vector representation we obtain
\[ (d Q Q^{-1})_{j} = -\frac12\varepsilon_{jik}(d Q Q^{-1})_{i k}
    = \varepsilon_{jik}(q_i d q_k - q_k d q_i)
    + \left(q_0 d q_r - q_r d q_0\right) \varepsilon_{jik}\varepsilon_{ikr}
\]
\[  = 2\varepsilon_{jik}q_i d q_k
    + 2\left(q_0 d q_r - q_r d q_0\right) \delta_{jr}
\]
\[  = 2\left(q_0 d q_j - q_j d q_0\right)
    + 2\varepsilon_{jik}q_i d q_k
\]

Hence was proved the relation (2), i.e.
\[ (d Q Q^{-1})_{i} = -\frac12\varepsilon_{irs}(d Q Q^{-1})_{rs}
    = 2\left(q_0 d q_i - q_i d q_0
    + \varepsilon_{irs}q_r d q_s\right) = 2 (d q q^{-1})_i
\]

$\blacksquare$

Let's call {\it arithmetic} vectors through bold symbols, i.e. sets of numbers that are vector components,
and let for angular velocity and intrinsic angular momentum they should be components in inertial frame of reference,
but for quaternions they should be in basis of generators of quaternion algebra.

Then in these coordinates
\[  d\Gamma_e = 2\cdot{\rm Id}
\leqno(6)\]
\[ \bsym{\omega} = 2 d\Gamma_e[\bsym{\xi}],\quad
   \bsym{\omega}\in Lie(SO(3)),\quad \bsym{\xi}\in Lie(S^3)
\leqno(7)\]

We have
\[
\begin{cases}
   d\Gamma_e[\bsym{\xi}] = \frac12 \bsym{\omega}; \\
   d\Gamma_e[\bsym{\xi}'] = \frac12 \bsym{\omega}'; \\
   d\Gamma_e[2\bsym{\xi}\times\bsym{\xi}'] = \bsym{\omega}\times\bsym{\omega}'; \\
\end{cases}
\leqno(8)\]

\newpage
\section{ Poisson structure for dynamic quaternion variables }
\label{Poisson2}

\bigskip
\subsection{ Jacobi identity for structure tensor }

Poisson structure for dynamic quaternion variables can be entered axiomatically.
\[
\begin{cases}
   \{q_\mu,q_\nu\} = 0, \quad \mu,\nu = 0,1,2,3; \\
   \{\mu_i,q_0\} = q_i ; \\
   \{\mu_i,q_j\} = \varepsilon_{ijk}q_k - q_0\delta_{ij};\\
   \{\mu_i, \mu_j\} = 2\varepsilon_{ijl}\mu_l;\\
\end{cases}
\leqno(1)\]

But for that we must prove the Jacobi identity for structure tensor of quaternionic Poisson brackets.

\bigskip
{\bf Proposition 4}. The structure tensor of quaternionic Poisson brackets (1) has form
\[
\begin{cases}
   J_{\mu\nu} = 0, \quad \mu,\nu = 0,1,2,3; \\
   J_{4+i,0} = q_i ; \\
   J_{4+i,j} = \varepsilon_{ijk}q_k - q_0\delta_{ij};\\
   J_{4+i,4+j} = 2\varepsilon_{ijl}\mu_l;\\
\end{cases}
\leqno(1a)\]

and Jacobi identity is satisfied
\[  J_{IJ}{}^{,L}J_{LK} + J_{JK}{}^{,L}J_{LI} + J_{KI}{}^{,L}J_{LJ} = 0
\leqno(2)\]

$\square$

If $I\leq 4$, $J\leq 4$ and $K\leq 4$ then execution of (2) is obvious.

Let's consider completely opposite case when $I>4$, $J>4$ and $K>4$. We have
\[  J_{i+4,j+4}{}^{,L}J_{L,k+4} + J_{j+4,k+4}{}^{,L}J_{L,i+4} + J_{k+4,i+4}{}^{,L}J_{L,j+4}
\]
\[ = (2\varepsilon_{ijr}\mu_r)^{,L}J_{L,k+4}
   + (2\varepsilon_{jkr}\mu_r)^{,L}J_{L,i+4}
   + (2\varepsilon_{kir}\mu_r)^{,L}J_{L,j+4}
\]
\[ = (2\varepsilon_{ijr}\mu_r)^{,l+4}J_{L,k+4}
   + (2\varepsilon_{jkr}\mu_r)^{,l+4}J_{L,i+4}
   + (2\varepsilon_{kir}\mu_r)^{,l+4}J_{L,j+4}
\]
\[ = (2\varepsilon_{ijr}\mu_r)^{,l+4}J_{l+4,k+4}
   + (2\varepsilon_{jkr}\mu_r)^{,l+4}J_{l+4,i+4}
   + (2\varepsilon_{kir}\mu_r)^{,l+4}J_{l+4,j+4}
\]
\[ = (2\varepsilon_{ijr}\delta_r^l)(2\varepsilon_{lks}\mu_s)
   + (2\varepsilon_{jkr}\delta_r^l)(2\varepsilon_{lis}\mu_s)
   + (2\varepsilon_{kir}\delta_r^l)(2\varepsilon_{ljs}\mu_s)
\]
\[ = 4(\varepsilon_{ijr}\varepsilon_{rks}
   + \varepsilon_{jkr}\varepsilon_{ris}
   + \varepsilon_{kir}\varepsilon_{rjs})\mu_s
\]
\[ = 4((\delta_{ik}\delta_{js} - \delta_{is}\delta_{jk})
   + (\delta_{ji}\delta_{ks} - \delta_{js}\delta_{ki})
   + (\delta_{kj}\delta_{is} - \delta_{ks}\delta_{ij}))\mu_s = 0
\]

If $I\leq 4$, $J\leq 4$ and $K>4$ then
\[  J_{\alpha\beta}{}^{,L}J_{L,k+4} + J_{\beta,k+4}{}^{,L}J_{L,\alpha} + J_{k+4,\alpha}{}^{,L}J_{L,\beta}
\]
\[  = J_{\beta,k+4}{}^{,L}J_{L,\alpha} + J_{k+4,\alpha}{}^{,L}J_{L,\beta}
    = - J_{k+4,\beta}{}^{,L}J_{L,\alpha} + J_{k+4,\alpha}{}^{,L}J_{L,\beta}
\]
\[  = - J_{k+4,\beta}{}^{,\gamma}J_{\gamma,\alpha} + J_{k+4,\alpha}{}^{,\gamma}J_{\gamma,\beta} = 0
\]

If now we have $I\leq 4$, $J>4$ and $K>4$ then
\[  J_{\alpha,j+4}{}^{,L}J_{L,k+4} + J_{j+4,k+4}{}^{,L}J_{L,\alpha} + J_{k+4,\alpha}{}^{,L}J_{L,j+4}
\leqno(3)\]
\[  = - J_{j+4,\alpha}{}^{,\gamma}J_{\gamma,k+4} + J_{j+4,k+4}{}^{,l+4}J_{l+4,\alpha} + J_{k+4,\alpha}{}^{,\gamma}J_{\gamma,j+4}
\]
\[  = J_{j+4,\alpha}{}^{,\gamma}J_{k+4,\gamma} - J_{k+4,\alpha}{}^{,\gamma}J_{j+4,\gamma}
    + J_{j+4,k+4}{}^{,l+4}J_{l+4,\alpha}
\]

And also let now $\alpha=0$ then expression (3) takes the form
\[  J_{j+4,0}{}^{,\gamma}J_{k+4,\gamma} - J_{k+4,0}{}^{,\gamma}J_{j+4,\gamma}
    + J_{j+4,k+4}{}^{,l+4}J_{l+4,0}
\]
\[  = J_{j+4,0}{}^{,s}J_{k+4,s} - J_{k+4,0}{}^{,s}J_{j+4,s}
    + 2\varepsilon_{ikl}q_l
\]
\[  = \delta^s_j (\varepsilon_{ksr}q_r - q_0\delta_{ks}) - \delta^s_k(\varepsilon_{jsr}q_r - q_0\delta_{js})
    + 2\varepsilon_{ikl}q_l
\]
\[  = (\varepsilon_{kjr}q_r - q_0\delta_{kj}) - (\varepsilon_{jkr}q_r - q_0\delta_{jk})
    + 2\varepsilon_{ikl}q_l
\]
\[  =  -2\varepsilon_{jkr}q_r + 2\varepsilon_{ikl}q_l = 0
\]

Else if $\alpha=t$ then expression (3) takes the form
\[  J_{j+4,t}{}^{,\gamma}J_{k+4,\gamma} - J_{k+4,t}{}^{,\gamma}J_{j+4,\gamma}
    + J_{j+4,k+4}{}^{,l+4}J_{l+4,t}
\]
\[  = J_{j+4,t}{}^{,0}J_{k+4,0} - J_{k+4,t}{}^{,0}J_{j+4,0}
    + 2\varepsilon_{jkl}J_{l+4,t}
    + J_{j+4,t}{}^{,s}J_{k+4,s} - J_{k+4,t}{}^{,s}J_{j+4,s}
\]
\[  = -\delta_{jt}q_k + \delta_{kt}q_j + 2\varepsilon_{jkl}(\varepsilon_{ltr}q_r - q_0\delta_{lt})
    + \varepsilon_{jts}(\varepsilon_{ksr}q_r - q_0\delta_{ks}) - \varepsilon_{kts}(\varepsilon_{jsr}q_r - q_0\delta_{js})
\]
\[  = -\delta_{jt}q_k + \delta_{kt}q_j + 2\varepsilon_{jkl}\varepsilon_{ltr}q_r
    + \varepsilon_{jts}\varepsilon_{ksr}q_r  - \varepsilon_{kts}\varepsilon_{jsr}q_r
\]
\[  - 2\varepsilon_{jkt}q_0 - \varepsilon_{jtk}q_0 + \varepsilon_{ktj}q_0
\]
\[  = -\delta_{jt}q_k + \delta_{kt}q_j + 2\varepsilon_{jkl}\varepsilon_{ltr}q_r
    + \varepsilon_{jts}\varepsilon_{ksr}q_r  - \varepsilon_{kts}\varepsilon_{jsr}q_r
\]
\[  = -\delta_{jt}q_k + \delta_{kt}q_j + 2(\delta_{jt}\delta_{kr} - \delta_{jr}\delta_{kt})q_r
    - (\delta_{jk}\delta_{tr} - \delta_{jr}\delta_{tk})q_r  + (\delta_{kj}\delta_{tr} - \delta_{kr}\delta_{tj})q_r
\]
\[  = -\delta_{jt}q_k + \delta_{kt}q_j + 2\delta_{jt}q_k - 2\delta_{kt}q_j
    - \delta_{jk}q_t + \delta_{tk}q_j  + \delta_{kj}q_t - \delta_{tj}q_k = 0
\]

Lists all possible combinations of indices.

$\blacksquare$

\bigskip
\subsection{ Poisson map from $S^3\times\mathbb{R}^3$ to $SO(3)\times\bsym{so(3)}^*$ }

Let's suppose now that on $S^3\times\mathbb{R}^3$ we have next Poisson structure
\[
\begin{cases}
   \{q_\mu,q_\nu\} = 0, \quad \mu,\nu = 0,1,2,3; \\
   \{\mu_i,q_0\} = q_i ; \\
   \{\mu_i,q_j\} = \varepsilon_{ijk}q_k - q_0\delta_{ij};\\
   \{\mu_i, \mu_j\} = 2\varepsilon_{ijl}\mu_l;\\
\end{cases}
\leqno(1)\]

Let's put in dynamic variables
\[ \pi_i = \frac12\mu_i
\leqno(2)\]

{\bf Proposition 5}. If Poisson brackets in quaternion variables (1) is valid then for matrices in form
\[ Q_{ik} = (2q_0^2 - 1)\delta_{ik}
    + 2 q_i q_k - 2 q_0 q_j\varepsilon_{jik}
\leqno(3)\]
the next Poisson brackets will fulfilled,
\[
\{Q_{ij}, Q_{kl}\} = 0, \quad
\{\pi_i, Q_{jk}\} = \varepsilon_{ijl}Q_{lk}, \quad
\{\pi_i, \pi_j\} = \varepsilon_{ijl}\pi_l
\leqno(4)\]
i.e. Poisson structure on $T^*SO(3)$ is induced from Poisson structure (1) on $T^*S^3$
put it in other way $\Gamma\times \frac12 \rm{id}_{R^3}$ is a Poisson mapping.

$\square$

From (1) follow next Poisson brackets
\[
\begin{cases}
   \{\pi_i,q_0^2\} = q_0 q_i; \\
   \{\pi_i,q_j q_k\} = \frac12\varepsilon_{ikn}q_j q_n + \frac12\varepsilon_{ijn}q_k q_n
                     - \frac12 q_0 q_j\delta_{ik} - \frac12 q_0 q_k\delta_{ij};\\
   \{\pi_i, q_0 q_l\} = -\frac12 q_0^2\delta_{il} + \frac12 q_i q_l + \frac12\varepsilon_{iln}q_0 q_n;\\
\end{cases}
\leqno(5)\]

Then
\[ \{\pi_i, Q_{jk}\} = 2\{\pi_i, q_0^2\}\delta_{jk} + 2\{\pi_i,q_j q_k\} - 2\varepsilon_{ljk}\{\pi_i, q_0 q_l\}
\]
\[ = 2 q_0 q_i\delta_{jk}
   + 2\left(\frac12\varepsilon_{ikn}q_j q_n + \frac12\varepsilon_{ijn}q_k q_n
   - \frac12 q_0 q_j\delta_{ik} - \frac12 q_0 q_k\delta_{ij}\right)
\]
\[ - 2\varepsilon_{ljk}\left(-\frac12 q_0^2\delta_{il} + \frac12 q_i q_l + \frac12\varepsilon_{iln}q_0 q_n\right)
\]
\[ = 2 q_0 q_i\delta_{jk}
   + \varepsilon_{ikn}q_j q_n + \varepsilon_{ijn}q_k q_n
   - q_0 q_j\delta_{ik} - q_0 q_k\delta_{ij}
\]
\[ + \varepsilon_{ljk} q_0^2\delta_{il} -\varepsilon_{ljk} q_i q_l - \varepsilon_{ljk}\varepsilon_{iln}q_0 q_n
\]
\[ = 2 q_0 q_i\delta_{jk}
   + \varepsilon_{ikn}q_j q_n + \varepsilon_{ijn}q_k q_n
   - q_0 q_j\delta_{ik} - q_0 q_k\delta_{ij}
\]
\[ + \varepsilon_{ljk} q_0^2\delta_{il} -\varepsilon_{ljk} q_i q_l + \varepsilon_{jlk}\varepsilon_{iln}q_0 q_n
\]
\[ = 2 q_0 q_i\delta_{jk}
   + \varepsilon_{ikn}q_j q_n + \varepsilon_{ijn}q_k q_n
   - q_0 q_j\delta_{ik} - q_0 q_k\delta_{ij}
\]
\[ + \varepsilon_{ljk} q_0^2\delta_{il} -\varepsilon_{ljk} q_i q_l
   + \left(\delta_{ij} \delta_{kn} - \delta_{ik} \delta_{jn}\right)q_0 q_n
\]
\[ = 2 q_0 q_i\delta_{jk}
   + \varepsilon_{ikn}q_j q_n + \varepsilon_{ijn}q_k q_n
   - q_0 q_j\delta_{ik} - q_0 q_k\delta_{ij}
\]
\[ + \varepsilon_{ljk} q_0^2\delta_{il} -\varepsilon_{ljk} q_i q_l
   + q_0 q_k \delta_{ij} - q_0 q_j\delta_{ik}
\]

Hence we have
\[ \{\pi_i, Q_{jk}\}
   = \varepsilon_{ljk} q_0^2\delta_{il} + 2 q_0 q_i\delta_{jk}
   - 2 q_0 q_j\delta_{ik}
\leqno(6)\]
\[ + \varepsilon_{ikn}q_j q_n + \varepsilon_{ijn}q_k q_n - \varepsilon_{ljk} q_i q_l
\]
\[ \{\pi_i, Q_{jk}\}
   = \varepsilon_{ljk} q_0^2\delta_{il} + 2 q_0 q_i\delta_{jk}
   - 2 q_0 q_j\delta_{ik}
\leqno(6a)\]
\[ + \varepsilon_{ikn}q_j q_n - \varepsilon_{jkn} q_i q_n  + \varepsilon_{ijn}q_k q_n
\]
\[ \{\pi_i, Q_{jk}\}
   = \varepsilon_{ljk} q_0^2\delta_{il} + 2 q_0 q_i\delta_{jk}
   - 2 q_0 q_j\delta_{ik}
\leqno(6b)\]
\[ + \left(\varepsilon_{ikn}\delta_{jl} - \varepsilon_{jkn}\delta_{il}\right)q_l q_n  + \varepsilon_{ijn}q_k q_n
\]

Since value $\varepsilon_{ikn}\delta_{jl} - \varepsilon_{jkn}\delta_{il}$
is antisymmetric with respect to $ij$ then it can be expressed by means of dual value.
\[ \frac12\varepsilon_{mij}\left(\varepsilon_{ikn}\delta_{jl} - \varepsilon_{jkn}\delta_{il}\right)
   = \delta_{kl} \delta_{mn} - \delta_{km} \delta_{nl}
\leqno(7)\]
i.e.
\[ \varepsilon_{ikn}\delta_{jl} - \varepsilon_{jkn}\delta_{il}
   = \varepsilon_{ijm}(\delta_{kl} \delta_{mn} - \delta_{km} \delta_{nl})
   = \varepsilon_{ijn}\delta_{kl} - \varepsilon_{ijk}\delta_{nl}
\leqno(7a)\]

Substituting (7a) in (6b) we obtain
\[ \{\pi_i, Q_{jk}\}
   = \varepsilon_{ljk} q_0^2\delta_{il} + 2 q_0 q_i\delta_{jk}
   - 2 q_0 q_j\delta_{ik}
\]
\[ + \left(\varepsilon_{ijn}\delta_{kl} - \varepsilon_{ijk}\delta_{nl}\right)q_l q_n  + \varepsilon_{ijn}q_k q_n
\]
\[ = \varepsilon_{ljk} q_0^2\delta_{il} + 2 q_0 q_i\delta_{jk}
   - 2 q_0 q_j\delta_{ik}
   - \varepsilon_{ijk}\bsym{q}^2 + 2\varepsilon_{ijn}q_k q_n
\]
\[ = \varepsilon_{ljk} q_0^2\delta_{il} + 2 q_0  q_n\left(\delta_{jk}\delta_{in} -  \delta_{jn}\delta_{ik}\right)
   - \varepsilon_{ijk}(1 - q_0^2) + 2\varepsilon_{ijn}q_k q_n
\]
\[ =  (2 q_0^2 - 1)\varepsilon_{ijk} + 2 q_0  q_n\left(\delta_{jk}\delta_{in} -  \delta_{jn}\delta_{ik}\right)
   + 2\varepsilon_{ijn}q_k q_n
\]
\[ =  (2 q_0^2 - 1)\varepsilon_{ijk} + 2 q_0  q_n\varepsilon_{ijm}\varepsilon_{nkm}
   + 2\varepsilon_{ijn}q_k q_n
\]
\[ =  \varepsilon_{ijl}\left((2 q_0^2 - 1)\delta_{lk}
   + 2 q_l q_k
   + 2 q_0  q_n\varepsilon_{nkl}\right)
\]
\[ =  \varepsilon_{ijl}\left((2 q_0^2 - 1)\delta_{lk}
   + 2 q_l q_k - 2 q_0 q_n\varepsilon_{nlk}\right)
\]

Now then
\[ \{\pi_i, Q_{jk}\}
   =  \varepsilon_{ijl}\left((2 q_0^2 - 1)\delta_{lk}
   + 2 q_l q_k - 2 q_0 q_n\varepsilon_{nlk}\right)
   = \varepsilon_{ijl}  Q_{lk}
\leqno(8)\]

$\blacksquare$

\newpage
\section{ Symplectic geometry on $T^*S^3\simeq S^3\times\mathbb{R}^3$ }
\label{Symplectic}

Let's consider Liouville-form on $T^*S^3$.
\[ \Theta = \mu_i(\delta q)^i,\quad \delta q = d q q^{-1} = d q q^\dag
\leqno(1)\]

It follows ($d(q^{-1})=-q^{-1}(d q)q^{-1}$)
\[ d\Theta = d\mu_i\w(\delta q)^i + \mu_i d(\delta q)^i
   = d\mu_i\w(\delta q)^i + \mu_i d(d q q^{-1})^i
\]
\[ = d\mu_i\w(\delta q)^i - \mu_i (d q\w d(q^{-1}))^i
   = d\mu_i(\delta q)^i + \mu_i (d q\w q^{-1}dqq^{-1})^i
\]

Let's consider 2-nd summand on pair of vectors
\[ d q\w q^{-1}dqq^{-1}(u,v)
   = d q(u)q^{-1}d q(v)q^{-1} - d q(v)q^{-1}d q(u)q^{-1}
\]
\[ = d qq^{-1}\w dqq^{-1}(u,v) = \delta q\w\delta q(u,v)
\]

Hence
\[ d\Theta = d\mu_i(\delta q)^i + \mu_i d(\delta q)^i
   = d\mu_i(\delta q)^i + \mu_i(\delta q\w\delta q)^i
\leqno(2)\]
\[ \Omega = -d\Theta
   = - d\mu_i\w(\delta q)^i - \mu_i(\delta q\w\delta q)^i
\leqno(2a)\]
\[ \Omega = (\delta q)^i\w d\mu_i - \mu_i(\delta q\w\delta q)^i
\leqno(2b)\]

Consider the expression $(\delta q\w\delta q)^i$ by calculating it on vectors $u,v$
\[ (\delta q\w\delta q)^i(u,v) = (\delta q(u)\delta q(v) - \delta q(v)\delta q(u))^i
\]
\[ = 2(\delta q(u)\times\delta q(v))^i = \varepsilon_{ikl}(\delta q^k(u)\delta q^l(v) - \delta q^k(v)\delta q^l(u))
\]
\[ = \varepsilon_{ikl}\delta q^k\w\delta q^l(u,v)
\]
i.e.
\[ (\delta q\w\delta q)^i = \varepsilon_{ikl}\delta q^k\w\delta q^l
\leqno(3)\]

Hence
\[ \Omega = \delta q^i\w d\mu_i - \mu_i\varepsilon_{ikl}\delta q^k\w\delta q^l
\leqno(2c)\]

\bigskip
\subsection{ Symplectic structure on $S^3\times\mathbb{R}^3$ that is induced from $SO(3)\times\bsym{so(3)}^*$ }
\[
\Omega^{T^*SO(3)}_{can}  = -d\Theta^{T^*SO(3)} = {\delta Q}^i\w d\pi_i - \pi_i [\delta Q,\delta Q]^i =
\leqno(1)\]
\[ = -\frac12\varepsilon_{ijk}\delta Q_{jk}\w d \pi_i
   + \frac12 \pi_i\varepsilon_{ijk}\delta Q_{js}\w \delta Q_{sk}
\]
Let's express differentials $\delta Q$ by means of $\delta q$
as well as using $d\pi = \frac12 d\mu$

We have
\[ -\frac12\varepsilon_{ijk}\delta Q_{jk}\w d \pi_i
   = \delta Q^i\w d\pi_i = 2\delta q^i\w\frac12 d\mu_i
   = \delta q^i\w d\mu_i
\]
and also
\[ \varepsilon_{ijk}\delta Q_{js}\w \delta Q_{sk}(u,v)
   = \varepsilon_{ijk}(\delta Q_{js}(u)\delta Q_{sk}(v) - \delta Q_{js}(v)\delta Q_{sk}(u))
\]
\[ = \varepsilon_{ijk}[\delta Q_{js}(u),\delta Q_{sk}(v)]
   = \varepsilon_{ijk}\widehat{(\delta \vec{Q}(u)\times\delta \vec{Q}(v))}_{jk}
   = -2(\delta \vec{Q}(u)\times\delta \vec{Q}(v))^i
\]
\[ = -8(\delta \vec{q}(u)\times\delta \vec{q}(v))^i
   = -8\varepsilon_{ijk}\delta q^j(u)\delta q^k(v)
\]
\[ = -4\varepsilon_{ijk}(\delta q^j(u)\delta q^k(v) - \delta q^j(v)\delta q^k(u))
   = -4\varepsilon_{ijk}(\delta q^j\w\delta q^k)(u,v)
\]
i.e.
\[ \varepsilon_{ijk}\delta Q_{js}\w \delta Q_{sk}(u,v)
   = -4\varepsilon_{ijk}(\delta q^j\w\delta q^k)(u,v)
\leqno(2)\]
or
\[ \varepsilon_{ijk}\delta Q_{js}\w \delta Q_{sk}
   = -4\varepsilon_{ijk}(\delta q^j\w\delta q^k)
\leqno(2a)\]

Thus
\[  \Omega^{T^*SO(3)}_{can}
    = -\frac12\varepsilon_{ijk}\delta Q_{jk}\w d \pi_i
    + \frac12 \pi_i\varepsilon_{ijk}\delta Q_{js}\w \delta Q_{sk}
\]
\[  = \delta q^i\w d\mu_i
    - 2\pi_i\varepsilon_{ijk}(\delta q^j\w\delta q^k)
     = \delta q^i\w d\mu_i
    - \mu_i\varepsilon_{ijk}(\delta q^j\w\delta q^k)
\]
that coincides with (2c) of previous subsection.

Then through direct calculation we proved in representation of right trivialization next proposition.

\bigskip
{\bf Proposition 6}. Symplectic structure on $T^*S^3$ is induced from symplectic structure on $T^*SO(3)$,
i.e.
\[
\begin{cases}
   \left(\Gamma\times \frac12 \rm{id}_{R^3}\right)^*\Theta^{T^*SO(3)}_{can} = \Theta^{T^*S^3}_{can};\\
   \left(\Gamma\times \frac12 \rm{id}_{R^3}\right)^*\Omega^{T^*SO(3)}_{can} = \Omega^{T^*S^3}_{can};\\
\end{cases}
\leqno(3)\]
and this mapping is revealed local isomorphism.

\bigskip
\subsection{ Left- and right-invariant field on $T^*S^3\simeq S^3\times~\mathbb{R}^3$ }

First of all, left and right action on $T^*S^3\simeq S^3\times\mathbb{R}^3$ takes the form
\[
\begin{cases}
   R_p(q,\bsym{\mu}) = (q p,\bsym{\mu});\\
   L_p(q,\bsym{\mu}) = (p q, p\bsym{\mu}p^{-1}) = (p q, \Gamma(p)[\bsym{\mu}]);\\
\end{cases}
\leqno(1)\]

Then
\[
\begin{cases}
   \tilde{\xi}^R(q,\bsym{\mu}) = (q\bsym{\xi},0);\\
   \tilde{\xi}^L(q,\bsym{\mu}) = (\bsym{\xi} q,\bsym{\xi}\bsym{\mu} - \bsym{\mu}\bsym{\xi})
   = (\bsym{\xi} q,2\bsym{\xi}\times\bsym{\mu});\\
\end{cases}
\leqno(2)\]

From (2) we have next proposition.

\bigskip
{\bf Proposition 7}. The vector fields
\[  \tilde{e_i}^L,\frac{\partial}{\partial\mu_i}
\leqno(3)\]
form the global basis on $T^*S^3\simeq S^3\times\mathbb{R}^3$.

Then an arbitrary vector
\[X\in T(S^3\times\mathbb{R}^3)\longrightarrow
  v^k\tilde{e}_k^L+\eta_k\frac{\partial}{\partial\mu_k}
\leqno(4)\]

\bigskip
\subsection{ Poisson structure of symplectic geometry on $T^*S^3$ }

We have
\[
\begin{cases}
   L_{\tilde{\xi}^R}\Theta = 0;\\
   L_{\tilde{\xi}^L}\Theta = 0;\\
\end{cases}
\leqno(1)\]

Then
\[  i_{\tilde{\xi}}\Omega = -i_{\tilde{\xi}} d\Theta
    = - L_{\tilde{\xi}}\Theta + d\circ i_{\tilde{\xi}}\Theta
    = d(\Theta(\tilde{\xi}))
\leqno(2)\]
and
\[
\begin{cases}
   i_{\tilde{\xi}^R}\Omega = d(\Theta(q\bsym{\xi},0))
   = d(\mu_i(q\bsym{\xi} q^{-1})^i) = d\langle\bsym{\mu},q\bsym{\xi}q^{-1}\rangle;\\
   i_{\tilde{\xi}^L}\Omega = d(\Theta(\bsym{\xi} q,2\bsym{\xi}\times\bsym{\mu}))
   = d(\mu_i\xi^i) = d\langle\bsym{\mu},\bsym{\xi}\rangle;\\
\end{cases}
\leqno(3)\]

2-nd line in (3) means that Hamiltonian field corresponding to dynamic variable
$\langle\bsym{\mu},\bsym{\xi}\rangle$ is $\tilde{\xi}^L$.

Then
\[ \{\langle\bsym{\mu},\bsym{\xi}\rangle, \langle\bsym{\mu},\bsym{\eta}\rangle\}
   = \partial_{\tilde{\eta}^L}\langle\bsym{\mu},\bsym{\xi}\rangle
   = 2\langle\bsym{\eta}\times\bsym{\mu},\bsym{\xi}\rangle
    = 2\langle\bsym{\mu},\bsym{\xi}\times\bsym{\eta}\rangle
\leqno(4)\]
\[ \{\bsym{\mu}, \langle\bsym{\mu},\bsym{\eta}\rangle\}
   = 2\bsym{\eta}\times\bsym{\mu}
\leqno(4a)\]
\[ \{\mu_i, \mu_k\} = 2\varepsilon_{ikl}\mu_l
\leqno(4b)\]

Analogically
\[ \{q^\alpha, \langle\bsym{\mu},\bsym{\xi}\rangle\}
   = \partial_{\tilde{\xi}^L}q^\alpha
   = (\bsym{\xi}q)^\alpha
\leqno(5)\]
or
\[ \{q, \langle\bsym{\mu},\bsym{\xi}\rangle\}
   = \bsym{\xi}q = - \langle\bsym{\xi},\bsym{q}\rangle e_0 + q^0\bsym{\xi} + \bsym{\xi}\times\bsym{q}
\leqno(5a)\]

Hence we have
\[
\begin{cases}
    \{q^0, \langle\bsym{\mu},\bsym{\xi}\rangle\} = - \langle\bsym{\xi},\bsym{q}\rangle;\\
    \{q^i, \langle\bsym{\mu},\bsym{\xi}\rangle\}
    = q^0\xi^i + (\bsym{\xi}\times\bsym{q})^i;\\
\end{cases}
\leqno(5b)\]
\[
\begin{cases}
    \{\langle\bsym{\mu},\bsym{\xi}\rangle, q^0\} = \langle\bsym{\xi},\bsym{q}\rangle;\\
    \{\langle\bsym{\mu},\bsym{\xi}\rangle, q^i\} = -q^0\xi^i - (\bsym{\xi}\times\bsym{q})^i;\\
\end{cases}
\leqno(5c)\]
\[
\begin{cases}
    \{\mu_i, q^0\} = q_i;\\
    \{\mu_i, q^k\} = -q^0\delta_i^k + \varepsilon_{ikl}q^l;\\
\end{cases}
\leqno(5d)\]

Now let's consider Poisson brackets for functions
that depend only on variables $q$ but not $\mu$.

For basis quaternions from (3) we obtain
\[ i_{\tilde{e}_k^L}\Omega = d\mu_k
\leqno(6)\]

Let coefficients $v^k$ are arbitrary functions $(q,\bsym{\mu})$.
Then we have
\[ i_{(v^k\tilde{e}_k^L)}\Omega = v^k d\mu_k
\leqno(6a)\]

Let
\[X=v^k\tilde{e}_k^L+\eta_k\frac{\partial}{\partial\mu_k} \in T(S^3\times\mathbb{R}^3)
\leqno(7)\]
--- arbitrary vector.

From (6a) and expression
\[ \Omega = (\delta q)^i\w d\mu_i - \mu_i\varepsilon_{ikl}\delta q^k\w\delta q^l
\leqno(8)\]
we have
\[ i_X\Omega = v^k d\mu_k + \eta_k\delta q^k
\leqno(9)\]

In equation for Hamiltonian field $X_F$ corresponding to Hamiltonian $F(q)$
\[ i_{X_F}\Omega = d F
\leqno(10)\]
as it is follows from (9)
the field $X_F$ can not contain components
related to basis vectors $\tilde{e}_k^L$
i.e.
\[X_F = \eta_k\frac{\partial}{\partial\mu_k}
\leqno(11)\]

If $F$ and $G$ are two such functions then from
\[  \{F, G\}(z) = \Omega(\xi_F(z),\xi_G(z))
\leqno(12)
\]
it should be
\[ \{F(q), G(q)\} = 0
\leqno(13)\]
and finally we have next proposition.

{\bf Proposition 8}. The Poisson structure of symplectic geometry on $T^*S^3$
is defined by next Poisson brackets
\[
\begin{cases}
   \{F(q), G(q)\} = 0;\\
   \{q, \langle\bsym{\mu},\bsym{\xi}\rangle\} = \bsym{\xi}q ;\\
   \{\langle\bsym{\mu},\bsym{\xi}\rangle, \langle\bsym{\mu},\bsym{\eta}\rangle\}
   = \langle\bsym{\mu},[\bsym{\xi},\bsym{\eta}]\rangle
   = 2\langle\bsym{\mu},\bsym{\xi}\times\bsym{\eta}\rangle;\\
\end{cases}
\leqno(14)\]

{\bf Remark 1}. Poisson brackets (14) are revealed coordinateless form
of Poisson brackets (1) \S8.

{\bf Remark 2}. So that get Hamiltonian field for variables $q_0,q_i$ from
\[ i_{X_q}\Omega = \eta_k\delta q^k
\leqno(15)\]
one can use the matrix of right action in the space of quaternions
\[ R_b =  \begin{bmatrix}
             b^0 &-b^1 &-b^2 &-b^3 \\
             b^1 & b^0 & b^3 &-b^2 \\
             b^2 &-b^3 & b^0 & b^1 \\
             b^3 & b^2 &-b^1 & b^0
          \end{bmatrix}
\leqno(16) \]

Then we get
\[ \bsym{\eta}^{(0)} = \begin{bmatrix}
                          -q^1\\ -q^2\\ -q^3 \\
                       \end{bmatrix}, \quad
   \bsym{\eta}^{(1)} = \begin{bmatrix}
                          q^0\\ q^3\\-q^2 \\
                       \end{bmatrix}, \quad
   \bsym{\eta}^{(2)} = \begin{bmatrix}
                          -q^3 \\ q^0 \\ q^1\\
                       \end{bmatrix}, \quad
   \bsym{\eta}^{(3)} = \begin{bmatrix}
                           q^2 \\ -q^1 \\ q^0\\
                       \end{bmatrix}
\leqno(17)\]

\newpage
\section{ Hamiltonian equations of motion for a rigid body in quaternion variables }
\label{MotionEq}

\bigskip
\subsection{ Poisson structure of a rigid body dynamics in quaternion variables in inertial frame of reference }

As shown in \cite{ZubHDLevOrb} right trivialization of cotangent bundle of $T^*SO(3)$
and $T^*S^3$ for description of a rigid body dynamics has the following advantage:
as far as translational and rotational degrees of freedom are separated
in inertial frame of reference;
then Poisson structure can be represented as direct product of structures 
that are refer to these degrees of freedom of rigid body.

Therefore, adding Poisson structure of subsection 8.1
by Poisson brackets for translational degrees of freedom we obtain
\[
\begin{cases}
   \{x_i,x_k\} = 0, \quad \{x_i,q_\mu\} = 0, \quad \{x_i,\mu_j\} = 0;\\
   \{p_i,q_\mu\} = 0, \quad \{p_i,\mu_j\} = 0;\\
   \{x_i,p_k\} = \delta_{ik}, \quad i, k = 0,1,2;\\
   \{q_\mu,q_\nu\} = 0, \quad \mu,\nu = 0,1,2,3; \\
   \{\mu_i,q_0\} = q_i ; \\
   \{\mu_i,q_j\} = \varepsilon_{ijk}q_k - q_0\delta_{ij};\\
   \{\mu_i, \mu_j\} = 2\varepsilon_{ijl}\mu_l;\\
\end{cases}
\leqno(1)\]

\bigskip
\subsection{ Poisson structure in quaternion variables in mixed frame of reference }

If we have Poisson brackets of previous subsection and system Hamiltonian
we can write Hamiltonian equations of motion.

But, as well known, expression of kinetic energy of a rigid body
is much more simpler in the body frame rather than in inertial frame of reference.

Therefore, it makes sense to conduct the canonical transformation
of generators of our Poisson structure,
namely intrinsic angular momentums of the body $\pi_i=\frac12\mu_i$
so that they represented to the body frame.

\[
\begin{cases}
   \bsym{\Pi} = \mathbf{Q}^{-1}[\bsym{\pi}]; \\
   \bsym{P} = \bsym{p},\quad \bsym{X} = \bsym{x},\quad \mathbf{Q} = \mathbf{Q}
\end{cases}
\leqno(1) \]

In quaternion variables 1-st line will be as follows
\[  {\bsym{\mathrm M}} = q^{-1}\bsym{\mu} q = Ad^\ast_{q}[\bsym{\mu}]
    = Q^{-1}[\bsym{\mu}]
\leqno(2)\]
or
\[ {\mathrm M}_j = Q_{rj}\mu_r
\leqno(2a)\]

That was revealed previously (see (8) in subsection 8.2)
\[ \{\mu_i,Q_{jk}\} = 2\varepsilon_{ijl}Q_{lk}
\leqno(3)\]

Therefore
\[ \{\mu_i, {\mathrm M}_j\} = \{\mu_i, Q_{rj}\mu_r\}
   =  Q_{rj}\{\mu_i,\mu_r\} + \{\mu_i, Q_{rj}\}\mu_r
\]
\[ =  2\varepsilon_{irt}Q_{rj}\mu_t + 2\varepsilon_{irt}Q_{tj}\mu_r
   =  2\varepsilon_{irt}Q_{rj}\mu_t + 2\varepsilon_{itr}Q_{rj}\mu_t
   = 0
\]
i.e.
\[ \{\mu_i, {\mathrm M}_j\} = 0
\leqno(4)\]

Consider also
\[  \{{\mathrm M}_i, Q_{jk}\} = \{Q_{ri}\mu_r, Q_{jk}\}
    = Q_{ri}\{\mu_r, Q_{jk}\} =  2\varepsilon_{rkt}Q_{ri}Q_{tj}
\]

Let's use identity
\[\varepsilon_{rst}Q_{ri}Q_{tj}Q_{sk} = \varepsilon_{ikj}
\leqno(5)\]
from it implies
\[\varepsilon_{rst}Q_{ri}Q_{tj} = Q_{kl}\varepsilon_{ilj}
\leqno(5a)\]

\newpage
\[  \{{\mathrm M}_i, Q_{jk}\} =  2\varepsilon_{rkt}Q_{ri}Q_{tj}
    = 2 Q_{kl}\varepsilon_{ilj} = -2 \varepsilon_{ijl}Q_{kl}
\]

Hence
\[  \{{\mathrm M}_i, Q_{jk}\} = -2 \varepsilon_{ijl}Q_{kl}
\leqno(6)\]

Now consider
\[ \{{\mathrm M}_i, {\mathrm M}_j\} = \{{\mathrm M}_i, Q_{rj}\mu_r\}
   = \{{\mathrm M}_i, Q_{rj}\}\mu_r + Q_{rj}\{{\mathrm M}_i, \mu_r\}
\]
\[ = \{{\mathrm M}_i, Q_{rj}\}\mu_r = -2 \varepsilon_{ijl}Q_{rl}\mu_r
   = -2 \varepsilon_{ijl}{\mathrm M}_l
\]
i.e.
\[ \{{\mathrm M}_i, {\mathrm M}_j\}  = -2 \varepsilon_{ijl}{\mathrm M}_l
\leqno(7)\]

Let's obtain Poisson brackets between moments and quaternion dynamic variables
\[  \{{\mathrm M}_i, q_0\}  = \{ Q_{ri}\mu_r, q_0\}
    =  Q_{ri}\{\mu_r, q_0\}
\]
\[  = Q_{ri}q_r  =
    \left((2q_0^2 - 1)\delta_{ri} + 2 q_r q_i - 2 q_0 q_l\varepsilon_{lri}\right)q_r
\]
\[  = (2q_0^2 - 1)q_i + 2\bsym{q}^2 q_i  = q_i
\]
i.e.
\[  \{{\mathrm M}_i, q_0\}  =  q_i
\leqno(8)\]

Now consider
\[  \{{\mathrm M}_i, q_j\}  = \{ Q_{ri}\mu_r, q_j\}
    = Q_{ri}\{\mu_r, q_j\}
\]
\[  = \left((2q_0^2 - 1)\delta_{ri} + 2 q_r q_i - 2 q_0 q_l\varepsilon_{lri}\right)
      \left(\varepsilon_{rjs}q_s - q_0\delta_{rj}\right)
\]
\[  = \left((2q_0^2 - 1)\delta_{ri} + 2 q_r q_i - 2 q_0 q_l\varepsilon_{lri}\right)\varepsilon_{rjs}q_s
\]
\[  - q_0\left((2q_0^2 - 1)\delta_{ri} + 2 q_r q_i - 2 q_0 q_l\varepsilon_{lri}\right)\delta_{rj}
\]
\[  = (2q_0^2 - 1)\varepsilon_{ijs}q_s
    - 2 q_0 q_l q_s\varepsilon_{lri}\varepsilon_{rjs}
\]
\[  - q_0(2q_0^2 - 1)\delta_{ij} - 2 q_0 q_j q_i + 2 q_0^2 q_l\varepsilon_{lji}
\]
\[  = (2q_0^2 - 1)\varepsilon_{ijl}q_l - 2 q_0^2 q_l\varepsilon_{ijl}
\]
\[  - q_0(2q_0^2 - 1)\delta_{ij} - 2 q_0 q_j q_i
\]
\[  + 2 q_0 q_l q_s( \delta_{jl}\delta_{is} - \delta_{ls}\delta_{ij} )
\]
\[  = -\varepsilon_{ijl}q_l - q_0(2q_0^2 - 1)\delta_{ij} - 2 q_0 q_j q_i
    + 2 q_0 q_j q_i - 2 q_0\bsym{q}^2\delta_{ij}
\]
\[  = -\varepsilon_{ijl}q_l - q_0(2q_0^2 - 1)\delta_{ij} - 2 q_0\bsym{q}^2\delta_{ij}
\]
\[  = -\varepsilon_{ijl}q_l - q_0(2q_0^2 + 2\bsym{q}^2 - 1)\delta_{ij}
    = -\varepsilon_{ijl}q_l - q_0\delta_{ij}
\]
i.e.
\[  \{{\mathrm M}_i, q_j\} = -q_0\delta_{ij} - \varepsilon_{ijl}q_l
\leqno(9)\]

Hence finally have next proposition.

{\bf Proposition 8}. The following Poisson structure in quaternion variables for rigid body dynamics
in mixed frame of reference is valid.

\[
\begin{cases}
   \{q_\mu,q_\nu\} = 0, \quad \mu,\nu = 0,1,2,3; \\
   \{{\mathrm M}_i, q_0\}  =  q_i;\\
   \{{\mathrm M}_i, q_j\} = -q_0\delta_{ij} - \varepsilon_{ijl}q_l;\\
   \{{\mathrm M}_i, {\mathrm M}_j\}  = -2 \varepsilon_{ijl}{\mathrm M}_l;\\
\end{cases}
\leqno(10)\]

\bigskip
\subsection{ Motion equations of a rigid body in mixed frame of reference }

Let's consider the Hamiltonian of rather general form
for Hamiltonian dynamics of a rigid body in quaternion variables.
\[ H\left((\bsym{x},\bsym{p}), (q, {\bsym{\mathrm M}})\right)
   = \frac1{2 m}\bsym{p}^2
   + T_{spin}\left((\bsym{x},\bsym{p}), (q, {\bsym{\mathrm M}})\right)
   + V(\bsym{x}, q)
\leqno(1)\]
where
\[ T_{spin}\left((\bsym{x},\bsym{p}), (q, {\bsym{\mathrm M}})\right)
   = \frac18 {\bsym{\mathrm M}} \mathbb{I}^{-1}{\bsym{\mathrm M}}
   = \frac18 \left(\frac{{\mathrm M}_1^2}{I_1} + \frac{{\mathrm M}_2^2}{I_2} + \frac{{\mathrm M}_3^2}{I_3}\right)
\leqno(2)\]

In mixed frame of reference arithmetic vectors $(\bsym{x},\bsym{p})$ 
(i.e. translational degrees of freedom) 
are components of the corresponding physical vectors in inertial frame of reference. 
And arithmetic vector $\bsym{\mathrm M}$ (rotational degree of freedom) 
are components of the corresponding physical vector in body frame; 
$\mathbb{I}$ --- constant tensor of inertia in the body frame.

\newpage
Using base Poisson brackets in mixed frame of reference
\[
\begin{cases}
   \{x_i,x_k\} = 0, \quad \{x_i,q_\mu\} = 0, \quad \{x_i,{\mathrm M}_j\} = 0;\\
   \{p_i,q_\mu\} = 0, \quad \{p_i,{\mathrm M}_j\} = 0;\\
   \{x_i,p_k\} = \delta_{ik}, \quad i, k = 0,1,2;\\
   \{q_\mu,q_\nu\} = 0, \quad \mu,\nu = 0,1,2,3; \\
   \{{\mathrm M}_i, q_0\}  =  q_i;\\
   \{{\mathrm M}_i, q_j\} = -q_0\delta_{ij} - \varepsilon_{ijl}q_l;\\
   \{{\mathrm M}_i, {\mathrm M}_j\}  = -2 \varepsilon_{ijl}{\mathrm M}_l;\\
\end{cases}
\leqno(3)\]
for the Hamiltonian (1) we obtain the next Poisson brackets
\[ \dot{x}_i = \{x_i,H\} = \left\{x_i,\frac1{2 m}\bsym{p}^2\right\} = \frac1{m}p_i
\]
\[ \dot{p}_i = \{p_i,H\} = \left\{p_i,V(\bsym{x}, q)\right\} = -\frac{\partial V}{\partial x_i}
\]

Stand for
\[ \Omega_i = \frac{\partial H}{\partial \pi_i}
   = 2\frac{\partial H}{\partial {\mathrm M}_i}
   = \frac12 (I^{-1})_{ik} {\mathrm M}_k
\leqno(4)\]
\[ \Omega_1 = \frac{{\mathrm M}_1}{2 I_1},\quad
   \Omega_2 = \frac{{\mathrm M}_2}{2 I_2},\quad
   \Omega_3 = \frac{{\mathrm M}_3}{2 I_3}
\leqno(4a)\]
\[ \dot{q}_0 = \{q_0,H\}
   = \frac{\partial H}{\partial {\mathrm M}_i}\left\{q_0,{\mathrm M}_i\right\}
\]
\[ = \frac12\Omega_i\left\{q_0,{\mathrm M}_i\right\}
   = - \frac12\Omega_i q_i
   = - \frac12\langle\bsym{\Omega}, \bsym{q}\rangle
\]
i.e.
\[ \dot{q}_0 = \{q_0,H\}
   = - \frac12\langle\bsym{\Omega}, \bsym{q}\rangle
\leqno(6)\]
in what follows
\[ \dot{q}_i = \{q_i,H\}
   = \frac{\partial H}{\partial {\mathrm M}_k}\left\{q_i,{\mathrm M}_k\right\}
\]
\[ = \frac12\Omega_k\left\{q_i,{\mathrm M}_k\right\}
   = \frac12\Omega_k\left( q_0\delta_{ki} + \varepsilon_{kil}q_l \right)
   = \frac12 q_0\Omega_i + \frac12\varepsilon_{kil}q_l\Omega_k
\]
\[ = \frac12 q_0\Omega_i - \frac12\varepsilon_{ikl}\Omega_k q_l
\]
i.e.
\[ \dot{q}_i = \{q_i,H\}
   = \frac12 q_0\Omega_i - \frac12\varepsilon_{ikl}\Omega_k q_l
\leqno(5)\]
or
\[ \dot{\bsym{q}} = \frac12 q_0\bsym{\Omega} - \frac12\bsym{\Omega}\times\bsym{q}
\leqno(6a)\]

in what follows
\[ \dot{{\mathrm M}}_i = \{{\mathrm M}_i,H\}
   = \frac{\partial H}{\partial q_0}\left\{{\mathrm M}_i, q_0\right\}
   + \frac{\partial H}{\partial q_k}\left\{{\mathrm M}_i, q_k\right\}
   + \frac{\partial H}{\partial {\mathrm M}_k}\left\{{\mathrm M}_i,{\mathrm M}_k\right\}
\]
\[ = \frac{\partial V}{\partial q_0}q_i
   + \frac{\partial V}{\partial q_k}\left( -q_0\delta_{ik} - \varepsilon_{ikl}q_l\right)
   + \frac12\Omega_k\left( -2 \varepsilon_{ikl}{\mathrm M}_l\right)
\]
\[ = \frac{\partial V}{\partial q_0}q_i
   - q_0\frac{\partial V}{\partial q_i}
   - \varepsilon_{ikl}\frac{\partial V}{\partial q_k} q_l
   - \varepsilon_{ikl}\Omega_k {\mathrm M}_l
\]
i.e.
\[ \dot{{\mathrm M}}_i
   = \frac{\partial V}{\partial q_0}q_i
   - q_0\frac{\partial V}{\partial q_i}
   - \varepsilon_{ikl}\frac{\partial V}{\partial q_k} q_l
   - \varepsilon_{ikl}\Omega_k {\mathrm M}_l
\leqno(7)\]
or
\[ \dot{{\bsym{\mathrm M}}}
   = \frac{\partial V}{\partial q_0}\bsym{q}
   - q_0\partial_{\bsym{q}}V
   - \partial_{\bsym{q}}V\times\bsym{q}
   - \bsym{\Omega}\times {\bsym{\mathrm M}}
\leqno(7a)\]

Now then
\[
\begin{cases}
   \dot{\bsym{x}} = \frac1{m}\bsym{p};\\

   \dot{\bsym{p}} = -\partial_{\bsym{x}}V;\\

   \dot{q}_0 = \{q_0,H\}
   = - \frac12\langle\bsym{\Omega}, \bsym{q}\rangle;\\

   \dot{\bsym{q}} = \frac12 q_0\bsym{\Omega} - \frac12\bsym{\Omega}\times\bsym{q};\\

   \dot{{\bsym{\mathrm M}}}
   = -\bsym{\Omega}\times {\bsym{\mathrm M}}
   + \frac{\partial V}{\partial q_0}\bsym{q}
   - q_0\partial_{\bsym{q}}V
   - \partial_{\bsym{q}}V\times\bsym{q};\\
\end{cases}
\leqno(8)\]

Let's lump together 4-th and 5-th line in (8).
\[  \dot{q} = - \frac12\langle\bsym{\Omega}, \bsym{q}\rangle e_0
    + \frac12 q_0\bsym{\Omega} - \frac12\bsym{\Omega}\times\bsym{q}
    = \frac12 q\bsym{\Omega}
\leqno(9)\]
\[  \dot{q} = \frac12 q\bsym{\Omega}
\leqno(9a)\]

Consider the quaternion
\[ \nabla^{(q)}V = e_0\partial_{q_0}V + \partial_{\bsym{q}}V
\leqno(10)\]

Then
\[ q^{-1}\nabla^{(q)}V = q^\dag\nabla^{(q)}V
   = \left(q_0 e_0 - \bsym{q}\right)\left(e_0\partial_{q_0}V + \partial_{\bsym{q}}V\right)
\]
\[  = \langle q, \nabla^{(q)}V\rangle e_0
    + q_0\partial_{\bsym{q}}V - \partial_{q_0}V\bsym{q} - \bsym{q}\times\partial_{\bsym{q}}V
\]
i.e.
\[  \partial_{q_0}V\bsym{q} - q_0\partial_{\bsym{q}}V + \bsym{q}\times\partial_{\bsym{q}}V
    = \langle q, \nabla^{(q)}V\rangle e_0  -  q^{-1}\nabla^{(q)}V
\]

Therefore
\[ \dot{{\bsym{\mathrm M}}}
   = -\bsym{\Omega}\times {\bsym{\mathrm M}}
   -  \left(q^{-1}\nabla^{(q)}V - \langle q, \nabla^{(q)}V\rangle\right)
\leqno(11)\]
and
\[
\begin{cases}
   \dot{\bsym{x}} = \frac1{m}\bsym{p};\\

   \dot{\bsym{p}} = -\partial_{\bsym{x}}V;\\

   \dot{q} = \frac12 q\bsym{\Omega}\longrightarrow q^{-1}\dot{q} = \frac12\bsym{\Omega};\\

  \dot{{\bsym{\mathrm M}}}
   = -\bsym{\Omega}\times {\bsym{\mathrm M}}
   -  \left(q^{-1}\nabla^{(q)}V - \langle q, \nabla^{(q)}V\rangle\right);\\
\end{cases}
\leqno(12)\]

\bigskip
{\bf Proposition 9}. The motion equations of a rigid body dynamics
for Hamiltonian~(1) in quaternion variables has the following algebraic form
\[
\begin{cases}
   \dot{\bsym{x}} = \frac1{m}\bsym{p};\\

   \dot{\bsym{p}} = -\partial_{\bsym{x}}V;\\

   \dot{q} = \frac12 q\bsym{\Omega}\longrightarrow q^{-1}\dot{q} = \frac12\bsym{\Omega};\\

  \dot{{\bsym{\mathrm M}}}
   = -\bsym{\Omega}\times {\bsym{\mathrm M}}
   -  \Im\left(q^{-1}\nabla^{(q)}V\right);\\
\end{cases}
\leqno(12a)\]
where $\Im(a)$ --- imaginary component of a quaternion $a$.

\end{document}